\documentclass[11pt]{article}
\usepackage[utf8]{inputenc}
\usepackage{calligra}
\usepackage[T1]{fontenc}
\usepackage{titling}
\usepackage{tabularx}
\usepackage{wrapfig}
\usepackage{lipsum}
\usepackage{amsmath}
\usepackage{empheq}
\usepackage[linktocpage,
colorlinks=true,
linkcolor=blue,
filecolor=red,      
urlcolor=red,
citecolor=red]{hyperref}
\usepackage{babel}
\usepackage{authblk}
\usepackage[margin = 1 in]{geometry}
\usepackage[dvipsnames]{xcolor}
\usepackage{graphicx}
\usepackage{float}
\usepackage{indentfirst}
\usepackage{caption}
\usepackage{subcaption}
\usepackage{wrapfig}
\newcommand{\tcr}{\textcolor{red}}

\newcommand{\rbm}[1]{{\color{blue}\bf [Robb: #1]}}

\newcommand{\brh}[2]{{\color{ForestGreen}\bf[Brayden: #1]}}

\numberwithin{equation}{section}
\title{\textbf{Thermodynamics of Exotic Black Holes 
\\ in Lovelock Gravity}}
\author[1,2,3]{Brayden Hull \thanks{b2hull@uwaterloo.ca}} 
\author[1,2,3]{Robert B. Mann 
\thanks{rbmann@uwaterloo.ca}}
\affil[1]{\textit{Department of Physics and Astronomy, University of Waterloo, Waterloo, Ontario, Canada, N2L 3G1} }
\affil[2]{\textit{Perimeter Institute for Theoretical Physics, 31 Caroline St. N., Waterloo, Ontario, N2L 2Y5, Canada}}
\affil[3]{\textit{Waterloo Center for Astrophysics, University of Waterloo, Waterloo ON, Canada, N2L 3G1}}
\date{}
\begin{document}
\begin{titlingpage}
    \maketitle
    \hrule
    \begin{abstract}
         We examine the thermodynamics of a new class of asymptotically AdS black holes with non-constant curvature event horizons in Gauss-Bonnet Lovelock gravity, with the cosmological constant acting as thermodynamic pressure. We find that non-trivial curvature on the horizon can significantly affect their thermodynamic behaviour.  We observe novel triple points in   6 dimensions between large and small uncharged black holes and thermal AdS.  For charged black holes we find a continuous set of triple points whose range depends on the parameters in the horizon geometry.  We also find new generalizations of massless and negative mass solutions previously observed in Einstein gravity.
    \end{abstract}

\end{titlingpage}
\newpage

\hrule 
\tableofcontents
\vspace{0.2 in}
\hrule

\newpage

\section{Introduction}
Einstein’s General Theory of Relativity is an extremely elegant and successful theory of gravitation, passing all of its experimental tests since its inception over 100 years ago \cite{Will_2014}. Yet its  reconciliation with quantum theory remains elusive.  While there is yet to be a proper full description of quantum gravity, a key piece of the puzzle is provided by black holes. Originally thought to be nature's ultimate repositories of matter and energy, their behaviour drastically changes once quantum physics is taken into account \cite{hawking1975particle}, leading to the well-known prediction that a   black hole   radiates like a thermal blackbody whose  temperature is proportional to its surface gravity and whose entropy is proportional to its horizon area.  In anti de Sitter spacetime they can undergo phase transitions \cite{Hawking:1982dh}, and in fact exhibit a very broad range of chemical thermodynamics \cite{Kubiznak:2016qmn}.  
   
Another general expectation that emerges from quantum gravity is the presence of higher curvature terms that correct the Einstein-Hilbert action \cite{birrell_davies_1982,PhysRevD.16.953}.   The most commonly discussed is Lovelock gravity \cite{Lovelock2,lovelock1971einstein}, which is regarded as a physically sensible generalization of Einstein gravity to higher dimensions since its field equations are second order in all metric components.  These theories have the general feature that the entropy of a black hole is no longer proportional to the horizon area, and so are of particular interest in black hole thermodynamics since they provide a window into how quantum gravitational effects could modify the radiative behaviour of black holes \cite{Frassino_2014}.

%While there has been considerable work on Gauss-Bonnet(2nd order) Lovelock gravity $CITE$ , there has been less work done on 3rd order lovelock.
%Such higher curvature theories no longer keep in the entropy simply proportional to the Area of the horizon, instead the introduction of higher curvature terms into the effective action yield the entropy proportional to the corrections due to these new terms. Thermodynamics aside these theories are of particular interest in the hunt for quantum gravity as there is understanding that when gravity becomes quantized there should be higher order terms in the action. \tcr{add sentence about why} .  

So far work on  black hole thermodynamics in Lovelock gravity
 has been limited to   black hole solutions that have a constant curvature manifold as its transverse space.  However this limitation is not necessary:
recently  a more general class of black hole solutions were found for  Lovelock gravity in which the transverse space is a more general manifold
\cite{Ray_2015}. Such solutions take the form of a warped product of a two-dimensional  space and an arbitrary transverse base manifold \cite{Dotti_2005,Dotti_2007,DOTTI_2009,Dotti_2010,Oliva_2013,Anabal_n_2011}. There is a generalization of the Birkhoff theorem that implies
this base manifold must be static. Furthermore, the field equations impose the conditions that 
 all the non-trivial intrinsic Lovelock tensors of the base manifold are constants that can be chosen arbitrarily.  We shall refer to such objects as 
`Exotic Lovelock Black Holes', or ELBHs. 

We investigate here the thermodynamic properties of  ELBHs in  Einstein Maxwell Gauss-Bonnet Gravity, the simplest Lovelock gravity theory. ELBHs in this case depend on two parameters, and in the limit that these parameters are chosen so that when the base manifold has constant curvature 
we recover     thermodynamic phenomena for both neutral and charged black holes previously observed in Gauss-Bonnet gravity \cite{Frassino_2014}. We find that ELBHs exhibit new effects as they increasingly depart from this special case. For example we observe a  novel triple point   between thermal radiation and large and small ELBHs in $6$ dimensions.  In the $d=6$ charged case we find that the particular features of the  large/intermediate/small triple point depends on the horizon geometry. 

We also find an interesting set of massless and negative mass black hole solutions that generalize those found previously in Einstein gravity \cite{Mann:1997jb,Aminneborg:1996iz,Smith:1997wx}.  These have a more interesting structure insofar as two horizons are possible under some circumstances, and their singularity structure is more complicated than the corresponding situation in Einstein gravity.

 We first begin with a review of Lovelock gravity, discussing black holes solutions whose transverse space is of constant and non-constant curvature and make a simple distinction between the two theories and how to transition between each theory, and provide a calculation of the Kretschmann scalar for an arbitrary metric. We follow that with Lovelock thermodynamics along with black hole thermodynamics. In Section 3 we examine Gauss-bonnet gravity, with first a discussion of vacumm solutions and then 5 and 6 dimensional black holes (uncharged then charged). In section 5 we give a summary of our results and future outlook.

\section{Exotic Lovelock Black Holes}

The Lagrangian for a Lovelock theory \cite{Lovelock2,lovelock1971einstein} in $d$ dimensions is  
\begin{equation} \label{Lag}
    \mathcal{L}=\frac{1}{16 \pi G_{N}} \sum_{k=0}^{K} \hat{\alpha}_{k} \mathcal{L}^{(k)}
\end{equation}
where $ \hat \alpha_{k}$ are the Lovelock coupling constants and $\mathcal{L}^{(k)}$ are the Euler densities 
\begin{equation}
    \mathcal{L}^{(k)}=\frac{1}{2^{k}} \delta_{c_{1} d_{1} \ldots c_{k} d_{k}}^{a_{1} b_{1} \ldots a_{k} b_{k}} R_{a_{1} b_{1}}^{c_{1} d_{1}} \ldots R_{a_{k} b_{k}}^{c_{k} d_{k}}.
\end{equation}
with the contraction occurring over the anti-symmetric generalized Kronecker delta.  The dimension of the Euler densities is $2k$, with  $\mathcal{L}^{(0)}$ beingthe cosmological constant, $\mathcal{L}^{(1)}$  the Ricci scalar, and $\mathcal{L}^{(2)}$ is the Gauss-Bonnett term. Note that we must have   $d> 2K$ in order to have non-trivial field equations.

With the lagragnian \eqref{Lag} we can write our action for Lovelock theory as
\begin{equation} \label{action}
    S= \int d^{d}x \sqrt{-g}\left( \frac{1}{16 \pi G_{N}}\sum_{k=0}^{K} \hat{\alpha}_{k} \mathcal{L}^{(k)} + \mathcal{L}_{matter} \right). 
\end{equation}
Variation of the action with respect to the metric yields the   field equations
\begin{equation} \label{vaceq}
    \sum_{k}^{K} \hat{\alpha}_{k} \mathcal{G}^{(k)}_{ ab}= 8\pi G_{N} T_{ab}
\end{equation}
where $T_{ab}$ is the stress-energy tensor and
$\mathcal{G}^{(k)}_{ab}$ are the Lovelock tensors
\begin{equation}
\mathcal{G}_{ ab}^{(k)}=-\frac{1}{2^{(k+1)}} g_{z a} \delta_{b e_{1} f_{1} \ldots e_{k} f_{k}}^{z c_{1} d_{1} \ldots c_{k} d_{k}} R_{c_{1} d_{1}}^{e_{1} f_{1}} \ldots R_{c_{k} d_{k}}^{e_{k} f_{k}}.
\end{equation}

We will be examing charged black holes which means we will use the following Lagrangian $\mathcal{L}_{matter} = -4 \pi G_{N} F_{a b} F^{a b}$, which will give us our finalized field equations
\begin{equation} \label{fieldeq}
\sum_{k=0}^{K} \hat{\alpha}_{(k)} \mathcal{G}_{a b}^{(k)}=8 \pi G_{N}\left(F_{a c} F_{b}^{c}-\frac{1}{4} g_{a b} F_{c d} F^{c d}\right).
\end{equation}
We shall discuss the black hole solutions to these equations in the next section.

\subsection{Black Hole Solutions}

We will be focusing on  charged AdS black hole solutions in this paper, using the ansatz
\begin{equation} \label{metric}
    ds^2 = 
{\textsf{g}}_{ij} {dy^i dy^j} 
    +  {\gamma}_{\alpha\beta}dx^\alpha dx^\beta
    =-f(r)dt^2 + \frac{dr^2}{f(r)} + r^2 d\Sigma_{d-2}^2 
    \qquad
    F = \frac{Q}{r^{d-2}}dt\wedge dr
\end{equation}
for the metric and gauge field strength,
 that we require to be  solutions to \eqref{fieldeq}. The coordinates $y^i = (t,r)$, with 
 ${\textsf{g}}_{ij} = \textrm{diag}(-f(r),1/f(r))$.
 The importance of this metric lies in the nature of the base manifold described by $d\Sigma_{d-2}^2$.

The most common approach is to take
$d\Sigma_{d-2}^2$ to be a $(d-2)$-dimensional compact space with constant curvature given by $(d-2)(d-3)\kappa$ with $\kappa=-1,0,+1$ corresponding to hyperbolic, flat, or spherical curvature respectively. This yields  the polynomial equation
\cite{PhysRevLett.55.2656,Cai_2004,Castro_2013,Camanho_2013,Takahashi_2012}
\begin{equation} \label{constpoly}
\sum_{k=0}^{K} \alpha_{k}\left(\frac{\kappa-f(r)}{r^{2}}\right)^{k}=\frac{16 \pi G M}{(d-2) \Sigma_{d-2}^{(\kappa)} r^{d-1}}-\frac{8 \pi G_{N} Q^{2}}{(d-2)(d-3)} \frac{1}{r^{2 d-4}}
\end{equation} 
from \eqref{fieldeq}.
Here M is the mass of the black hole, $\Sigma_{d-2}^{(\kappa)}$ is the volume of the compact space whose metric is 
$d\Sigma_{d-2}^2$. 
Q is the black hole charge given by 
\begin{equation}
    Q=\frac{1}{2 \Sigma_{d-2}^{(\kappa)}} \int * F 
\end{equation}

%\brh{Sould the Q on the equation with the wedge product be little q?}

The $\alpha_k$ terms are re-scaled Lovelock coupling constants 
\begin{equation} 
\alpha_{0}=\frac{\hat{\alpha}_{(0)}}{(d-1)(d-2)}, 
\quad \alpha_{1}=\hat{\alpha}_{(1)}, \quad \alpha_{k}=\hat{\alpha}_{(k)} \prod_{n=3}^{2 k}(d-n) \quad \text { for } \quad k \geq 2
\end{equation}
where the cosmological constant
$\Lambda = -\hat{\alpha}_{(0)}/2$. In what follows we will set $\alpha_{1}=1$ to retrieve general relativity in the low energy limit.

However it is possible to 
assume that
$d\Sigma_{d-2}^2$ is the metric of a more general $(d-2)$-dimensional base manifold that does not have to be of constant curvature. In this case the field equations   \eqref{fieldeq} become
\begin{equation}
  \mathcal{G}^{i}_{j} \equiv -\frac{ (d-2)(d-1) \delta_{j}^{i}}{2(d-1) ! r^{d-2}} \sum_{n=0}^{\bar{k}}\left\{(d-2 n-2) ! \hat{\mathcal{L}}^{(n)}\right\}\left\{\frac{d}{d r}\left(r^{d-2 n-1} A_{n}\left(\frac{-f(r)}{r^2}\right )\right)\right\}= 8 \pi G_{N} T^{i}_{j} \label{rawfield1}
\end{equation}  
\begin{equation}
 \mathcal{G}^{\alpha}_{\beta} \equiv \left.\frac{(d-1)(d-2)}{(d-1) ! r^{d-3}} \sum_{n=0}^{\bar{k}}\left\{(d-2 n-3) ! \hat{\mathcal{G}}^{(n)\alpha}_{\beta}\right\}\right\}\left\{\frac{d^{2}}{d r^{2}}\left(r^{d-2 n-1} A_{n}\left(\frac{-f(r)}{r^2}\right)\right)\right\} = 8 \pi G_{N} T^{\alpha}_{\beta} \label{rawfield2}
\end{equation}
with the polynomial 
\begin{equation}
    A_{n}\left(\frac{-f(r)}{r^{2}}\right) \equiv \sum_{k=n}^{K} \alpha_{k} \binom{k}{n} \left( \frac{-f(r)}{r^{2}}\right)^{k-n} 
\end{equation}
which satisfies the recurrence relation
\begin{equation}
    A_{n}^{\prime}\left(\frac{-f(r)}{r^{2}}\right)=(n+1)A_{n+1}\left(\frac{-f(r)}{r^{2}}\right).
\end{equation}

The quantities $\hat{\mathcal{L}}^{(n)}$ and $\hat{\mathcal{G}}^{(n)\alpha}_{\beta}$ are respectively the Euler Characteristic and Lovelock Tensors of the base manifold
\begin{equation}
    \hat{\mathcal{L}}^{(n)}=\frac{(d-2)! b_{n}}{(d-2 n-2) !} \qquad \qquad \hat{\mathcal{G}}^{(n)\alpha}_{\beta}=-\frac{(d-3)!b_{n}}{2 (d-2n-3)!}\delta^{\alpha}_{\beta}
\label{Eul-Lov}
\end{equation}
reducing the field equations to
\begin{equation}
    \mathcal{G}_{j}^{i} \equiv \frac{-(d-2) \delta_{j}^{i}}{2 r^{d-2}} \frac{d}{d r} \sum_{n=0}^{K}\left\{b_{n}\left(r^{d-2 n-1} A_{n}\left(\frac{-f(r)}{r^{2}}\right)\right)\right\}=8 \pi G_{N} T_{j}^{i} \label{finalfield1}
\end{equation}
\begin{equation}
    \mathcal{G}^{\alpha}_{\beta} \equiv \frac{- \delta^{\alpha}_{\beta}}{2 r^{d-3}} \frac{d^2}{dr^2} \sum_{n=0}^{K} \left\{ b_{n}\left( r^{d-2 n-1} A_{n}\left(\frac{-f(r)}{r^{2}}\right) \right) \right\} = 8 \pi G_{N} T^{\alpha}_{\beta} \label{finalfield2}
\end{equation}
In this case the polynomial equation in $f(r)$ becomes \cite{Ray:2015ava}
\begin{equation} \label{souryapoly}
    \sum_{n=0}^{K} \frac{b_{n}}{r^{2n}} \left( \sum_{k=n}^{K} \alpha_{k} \binom{k}{n}\left(\frac{-f(r)}{r^{2}}\right)^{k-n}\right )=\frac{16 \pi G_{N} M}{(d-2) \Sigma_{d-2} r^{d-1}}-\frac{8 \pi G_{N} Q^{2}}{(d-2)(d-3)} \frac{1}{r^{2 d-4}}
\end{equation}
generalizing \eqref{constpoly}.  
We shall refer to the solutions that follow from solutions 
to \eqref{souryapoly} in which $f(r)$ vanishes at least once for some $r>0$ as
\textit{Exotic Lovelock Black Holes}, or ELBHs, where the term $b_{n}$ is introduced and is referred to as the topological parameter. We can set $b_{0}=1$ without any loss of generality.  

Before proceeding to the thermodynamics we   mention a few more things with regards to the  field equations \eqref{finalfield1} and \eqref{finalfield2}.   Upon comparing the left-hand sides of equations \eqref{constpoly} and \eqref{souryapoly}, we obtain
\begin{equation}\label{bncc}
    b_{n}=\kappa^n
\end{equation}
for black holes whose base manifolds have constant curvature. 

We define the mass, using the Hamiltonian formulation, as the conserved charge corresponding to the time translational Killing vector of a background spacetime to which the black hole solutions approach in the asymptotic region. For black holes with maximally symmetric horizons, this is usually chosen as a constant curvature spacetime that solves the field equations \cite{Frassino_2014,Cai:2003kt,2013JHEP...07..164C,Camanho:2011rj,Takahashi:2011du}. In our case, we choose this background solution to have identical geometry of the base manifold as that of the black hole solution under consideration while the corresponding metric function $\bar{f}(r)$ solves the equation
\begin{equation} \label{souryapolyref}
    \sum_{n=0}^{K} \frac{b_{n}}{r^{2n}} \left( \sum_{k=n}^{K} \alpha_{k} \binom{k}{n}\left(\frac{-\bar{f}(r)}{r^{2}}\right)^{k-n}\right )=0
\end{equation}

One might note an apparent discrepancy in the above equation while considering the dimension $d=2K+1$, since  in this case there are $K-1$ constants $b_n$ characterizing the geometry of the base manifold. In this particular dimensionality, $b_K$ does not carry any geometric information of the base manifold but rather corresponds to the choice of the integration constant specifying the background spacetime.
  This becomes evident if we rewrite the above equation as
\begin{equation} \label{souryapolyref2}
    \sum_{n=0}^{K-1} \frac{b_{n}}{r^{2n}} \left( \sum_{k=n}^{K} \alpha_{k} \binom{k}{n}\left(\frac{-\bar{f}(r)}{r^{2}}\right)^{k-n}\right )=-\frac{b_K\alpha_K}{r^{2K}}
\end{equation}
For spherical base manifolds, $b_K=1$ then corresponds to the natural choice of a constant curvature spacetime as the background. We shall adopt the convention that the sum in \eqref{souryapoly} extends up to $K$ for all $d$, recognizing that if $d=2K+1$ the parameter $b_K$ corresponds to a convention for choosing $M$ and contains no geometric information.

In analyzing the structure of our solutions, we will find it useful to employ the  Kretschmann scalar, which for the metric \eqref{metric} can be written as
\begin{equation}\label{Kscal}
    R^{ abcd }R_{abcd }=\left(\frac{d^{2} f(r)}{d r^2}\right)^{2} +  2\frac{(d-2)}{r^2}\left( \frac{d f(r)}{dr} \right) ^2 + 2\frac{(d-2)(d-3) f(r)^2}{r^4}  - 4\frac{ R[\gamma] f(r)}{r^4} +\frac{\mathcal{K}[\gamma]}{r^4}
\end{equation}
where 
\begin{equation}
     \mathcal{K}[\gamma]=R^{ \alpha \beta \mu \nu }R_{ \alpha \beta \mu \nu }[\gamma]
\end{equation}
is the Kretschmann scalar of the base manifold.

\subsection{Lovelock Black Hole Thermodynamics}

Lovelock black holes of mass $M$, entropy $S$, temperature $T$, and charge $Q$   obey the extended First Law and Smarr relations \cite{Jacobson_1993,Kastor_2010}  
\begin{equation}
\delta M=T \delta S-\frac{1}{16 \pi G_{N}} \sum_{h} \hat{\Psi}^{(k)} \delta \hat{\alpha}_{(k)}+\Phi \delta Q
\end{equation}
\begin{equation}
(d-3) M=(d-2) T S+\sum_{k} 2(k-1) \frac{\hat{\Psi}^{(k)} \hat{\alpha}_{(k)}}{16 \pi G_{N}}+(d-3) \Phi Q
\end{equation}
where we regard the $\hat \alpha_{(k)}$ as thermodynamic parameters. The $\hat \Psi^{(k)}$ are their respective conjugate thermodynamic potentials, given by
\begin{equation}
\hat{\Psi}^{(k)}=4 \pi T \mathcal{A}^{(k)}+\mathcal{B}^{(k)}+\Theta^{(k)}
\end{equation}
with 
\begin{equation}
\begin{aligned}
\mathcal{B}^{(k)} &=-\frac{16 \pi k G_{N} M(d-1) !}{b(d-2 k-1) !}\left(-\frac{1}{\ell^{2}}\right)^{k-1}, \quad b=\sum_{k} \frac{\hat{\alpha}_{k} k(d-1) !}{(d-2 k-1) !}\left(-\frac{1}{\ell^{2}}\right)^{k-1} \\
\Theta^{(k)} &=\int_{\Sigma} \sqrt{-g} \mathcal{L}^{(k)}[s]-\int_{\Sigma_{\mathrm{AdS}}} \sqrt{-g_{\mathrm{AdS}}} \mathcal{L}^{(k)}\left[s_{\mathrm{AdS}}\right]
\end{aligned}
\end{equation}
where $ \ell^2 = 1/\alpha^{(0)}$ is the 'AdS' radius. The spatial hypersurface $\Sigma$,
with timelike unit normal $n^a$ and induced metric $s_{a b}=g_{a b}+n_{a} n_{b}$, 
extends from the horizon to infinity.

Black holes in Lovelock gravity no longer obey the area relation $S=\frac{A_{H}}{4}$, but instead have entropy given by \cite{Jacobson_1993}
\begin{equation}
S=\frac{1}{4 G_{N}} \sum_{k} \hat{\alpha}_{k} \mathcal{A}^{(k)}, \quad \mathcal{A}^{(k)}=k \int_{\mathcal{H}} \sqrt{\sigma} \mathcal{L}^{(k-1)}
\end{equation}
where $\sigma$ is the determinant of the  induced metric on the horizon, and $\mathcal{L}^{(k-1)}$ are the corresponding Euler densities. 

We will be treating the (negative) cosmological constant  as the thermodynamic pressure
\begin{equation} \label{pressure}
P=-\frac{\Lambda}{8 \pi G_{N}}=\frac{\hat{\alpha}_{0}}{16 \pi G_{N}}, \quad V=-\hat{\Psi}^{(0)}
\end{equation}
with $V$ the conjugate thermodynamic volume of the black hole.

\subsubsection{ELBH Thermodynamics}

We wish to solve \eqref{metric}  to obtain AdS black hole solutions, solutions for which  $f(r)$ grows quadratically with $r$ for large $r$ and has $r=r_+ >0$ as its largest linear zero.  Although we could explicitly solve \eqref{metric} in the $K=2$ Gauss-Bonnet case, this is not necessary as we can employ the Hamiltonian formalism 
 \cite{Cai_2004,Kastor_2010,RayExotic}. We can find the thermodynamic parameters of the black hole without an explicit solution of $f(r)$ for any value of $K$.
 
Setting $f(r_+)=0$, we find 
\begin{equation}
    M=\frac{\Sigma_{d-2}(d-2)}{16 \pi G_N} \sum_{k=0}^{K} \alpha_{k} b_{k} r_{+}^{d-1-2 k}+\frac{\Sigma_{d-2}}{2(d-3)} \frac{Q^{2}}{r_{+}^{d-3}}.
\end{equation}
\begin{equation} \label{temp}
T=\frac{f^{\prime}\left(r_{+}\right)}{4 \pi}=\frac{1}{4 \pi r_{+} D\left(r_{+}\right)}\left[\sum_{k=0}^{K} b_{k} \alpha_{k}(d-2 k-1)r_{+}^{-2(k-1)}-\frac{8 \pi G_N Q^{2}}{(d-2) r_{+}^{2(d-3)}}\right]
\end{equation}
where $D(r_{+})$ is
\begin{equation}
D\left(r_{+}\right)=\sum_{k=1}^{K} k \alpha_{k}b_{k-1} r_{+}^{-2(k-1)}\; .
\end{equation}
The entropy is given by
\begin{equation}
S=\frac{\Sigma_{d-2}(d-2)}{4 G_N} \sum_{k=0}^{K} \frac{k b_{k-1} \alpha_{k} r_{+}^{d-2 k}}{d-2 k}
\end{equation}
and through the first law it is easy to identify the conjugate potentials 
\begin{equation}
\Psi^{(k)}=\frac{\Sigma_{d-2}(d-2)}{16 \pi G_{N}}  r_{+}^{d-2 k}\left[\frac{b_{k}}{r_{+}}-\frac{4 b_{k-1} \pi k T}{d-2 k}\right].
\end{equation}
The thermodynamic volume is 
\begin{equation}\label{volume}
V=-\hat{\Psi}^{(0)}=\frac{16 \pi G_{N} \Psi^{(0)}}{(d-1)(d-2)}=\frac{\Sigma_{d-2}r_{+}^{d-1}}{ {d-1}}.
\end{equation}
Using \eqref{pressure} and \eqref{temp} we obtain the equation of state
\begin{equation}\label{eos}
P=\frac{d-2}{16 \pi G_N} \sum_{k=1}^{K} \frac{\alpha_{k}}{r_{+}^{2}}r_{+}^{-2(k-1)}\left[4 \pi k b_{k-1} r_{+} T-b_{k}(d-2 k-1)\right]+\frac{Q^{2}}{2  r_{+}^{2(d-2)}}
\end{equation}
where $r_+$ is a function of the thermodynamic volume from \eqref{volume}. 

We note the non-trivial dependence of the various thermodynamic parameters on the topological constants $b_k$. This implies that we might expect new phase behaviour for ELBHs as compared to their constant curvature counterparts.  To investigate this we shall study the  Gibbs free energy \cite{Kubiznak:2016qmn}
\begin{equation}\label{gibbs0}
    G(P,T,Q) =M-TS
\end{equation}
which characterizes the canonical ensemble.  A thermodynamically stable state is given by the global minimum of $G$ for any given choice of the parameters. To observe phase transitions, it is most useful to plot $G$ as a function of $T$, fixing the other parameters.  

This, however, is not sufficient to determine a physically acceptable black hole thermodynamic state.  We shall also require that 
\begin{equation}
    \sum_{k=0}^{K} \frac{k b_{k-1} \alpha_{k} r_{+}^{d-2 k}}{d-2 k} \geq 0. \label{posentropy}
\end{equation}
so that the entropy \label{entropy} is not negative. Likewise, we shall only consider $T\geq 0$ in \eqref{temp}.  However we shall not require that $M>0$, since it is known that, for example, topological black holes can have negative masses that are bounded from below \cite{Mann:1997jb}.

\section{ELBHs in Gauss-Bonnet Gravity}

We now specialize our considerations to the $K=2$ Gauss-Bonnet case.

\subsection{Solutions}

\noindent Setting $K=2$, in \eqref{souryapoly} we obtain at the polynomial equation for $f(r) \equiv f$ in Gauss-Bonnet-Lovelock gravity (where we recall $b_{0}=1$  and $\alpha_{1}=1$)
\begin{equation}
\frac{\alpha_{2} f^{2}}{r^{4}}+\left(-\frac{1}{r^{2}}-\frac{2 b_{1} \alpha_{2}}{r^{4}}\right) f+\alpha_{0}+\frac{b_{1}}{r^{2}}+\frac{b_{2} \alpha_{2}}{r^{4}}=\frac{16 \pi M}{(d-2) \Sigma_{d-2} r^{d-1}}-\frac{8 \pi Q^{2}}{(d-2)(d-3) r^{2 d-4}}
\end{equation}
whose solutions are
\begin{equation}\label{elbhsol}
f= f_\pm(\textsf{m},\textsf{q}) \equiv \frac{r^{2}+2 b_{1} \alpha_{2} \pm \sqrt{\left( b_{1}^{2}- b_{2}\right)4 \alpha_{2}^{2} + r^4(1-4 \alpha_{2} \alpha_{0})
+\frac{8\textsf{m}\alpha_{2}}{r^{d-5}}
-\frac{4 \textsf{q}^2 \alpha_{2}}{r^{2(d-4)}}
}}{2 \alpha_{2}}
\end{equation}
where
\begin{equation}
\textsf{m} \equiv \frac{8 \pi M}{(d-2) \Sigma_{d-2}}   \qquad
\textsf{q}^2 \equiv \frac{8 \pi Q^{2}}{(d-2)(d-3)}\; .
\end{equation}
The solution $f_{-}(\textsf{m},\textsf{q})$ has the limit
\begin{equation}\label{einlim}
\lim_{\alpha_2 \to 0} f_{-}(\textsf{m},\textsf{q}) = \alpha_0 r^2 + b_1 - \frac{2\textsf{m}}{r^{d-3}} + \frac{\textsf{q}^2}{r^{2(d-3)}}
\end{equation}
recovering the solution in Einstein  gravity for $b_1 =\kappa$.

We also require that $f_\pm(\textsf{m},\textsf{q})\to r^2$ for large $r$. This implies
\begin{equation}\label{largercond}
    1 - 4 \alpha_{2} \alpha_{0}\geq 0.
\end{equation}
independent of the $b_k$.  Equation
\eqref{largercond} implies from \eqref{pressure} that there is a maximum pressure \cite{Frassino_2014}
\begin{equation}\label{maxP}
P \leq  P_{\max}= \frac{(d-1)(d-2)}{64 \pi \alpha_2}  
\end{equation}
such that if this bound is violated the spacetime is no longer asymptotically AdS.

The horizons are located at 
\begin{equation}\label{horpm}
r_{\pm}^2(\textsf{m},\textsf{q}) = -\frac{1}{2\left(\alpha_0-\frac{2\textsf{m}}{r_\pm^{d-5}}
-\frac{\textsf{q}^2}{r_\pm^{2(d-4)}} \right)}\left(
b_1 \pm \sqrt{b_1^2 - 4 b_2 \alpha_2\alpha_0
+b_2 \alpha_2\left(\frac{8\textsf{m}}{r_\pm^{d-5}}
-\frac{4\textsf{q}^2}{r_\pm^{2(d-4)}} \right)
}
\right)    
\end{equation}
which implicitly defines $r_\pm$.  We note from this that solutions 
with $\textsf{m}=\textsf{q}=0$
\begin{equation}\label{vacsol}
f_\pm(0,0) =  \frac{r^{2}+2 b_{1} \alpha_{2} \pm \sqrt{\left( b_{1}^{2}- b_{2}\right)4 \alpha_{2}^{2} + r^4(1-4 \alpha_{2} \alpha_{0})}}{2 \alpha_{2}}
\end{equation}
have horizons at
\begin{equation}\label{vachor}
r_{\pm}^2(0,0) = \frac{1}{2\alpha_0}\left(-
b_1 \pm \sqrt{b_1^2 - 4 b_2 \alpha_2\alpha_0}
\right)    
\end{equation}
provided either 
\begin{align}\label{cond1}
\textrm{(a)}&\quad  b_2<0 \quad\textrm{and}\quad b_1>0\quad
\Rightarrow   r_{+}(0,0) \;\textrm{is the only horizon} \\
& \nonumber
\\
\textrm{or, if}\quad   
 b_2 > 0 \nonumber\\ 
\textrm{(b)}&\quad  b_1^2 > 4  b_2 \alpha_2 \alpha_0 \quad \textrm{and}\quad 0 >  b_1 > 
- \sqrt{\frac{b_2 }{2(1-2\alpha_2\alpha_0)}}
\quad
\Rightarrow  r_{\pm}(0,0) \;\textrm{are both horizons}
\label{cond2}
\\ 
\textrm{or}
\nonumber\\ 
\textrm{(c)}&\quad  b_1^2 > 4  b_2 \alpha_2 \alpha_0 \quad \textrm{and}\quad  - {\sqrt{\frac{ b_2 }{2(1-2\alpha_2\alpha_0)}}} > b_1  \quad \Rightarrow r_{+}(0,0) \;\textrm{is the only horizon}
\label{cond3}
\end{align}
where the   inequalities \eqref{cond2} ensures that $r_{-}(0,0)$ is real and larger than the location of the spacetime singularity.

 If any of these conditions do not hold then the solution has a naked singularity.

The solutions \eqref{vacsol} are generalizations of massless topological black holes
in Einstein gravity
\cite{Mann:1997jb,Aminneborg:1996iz,Smith:1997wx}, with $\alpha_2=0$ and $b_1=-1$, with appropriate identifications made on the transverse base space  \cite{Aminneborg:1996iz,Mann:1997iz}. Here we have a richer set of possibilities insofar as two horizons are possible, as long as $b_2\alpha_2 > 0$.  Negative mass solutions are likewise possible.

An evaluation of the  Kretschmann Scalar \eqref{Kscal} for the 
solution \eqref{vacsol} yields
\begin{align}
 R^{ abcd }R_{abcd } &=\frac{1}{\alpha_2}+
{\frac {-48\, \left( {{ b_1}}^{2}-{ b_2} \right) {r}^{2}{ \alpha_0}\,
{{ \alpha_2}}^{3}+12\,{r}^{2} \left( 4{r}^{4}{{ \alpha_0}}^{2}/3+{{ b_1}
}^{2}-{ b_2} \right) {{ \alpha_2^2}}-8\,{ \alpha_0}\,{ \alpha_2}\,{r}^{6}}{{ \alpha_2}\,
 \left(\left( b_{1}^{2}- b_{2}\right)4 \alpha_{2}^{2} + r^4(1-4 \alpha_{2} \alpha_{0}) \right) ^{3/2}}}
\nonumber\\
&-2\,{\frac {
 \left( d-2 \right)  \left( -4\,{ \alpha_0}\,{ \alpha_2}\,{r}^{2}+{r}^{2}+
\sqrt { \left( b_{1}^{2}- b_{2}\right)4 \alpha_{2}^{2} + r^4(1-4 \alpha_{2} \alpha_{0})} \right) ^{2}}{{{ \alpha_2^2}}
 \left( \left( b_{1}^{2}- b_{2}\right)4 \alpha_{2}^{2} + r^4(1-4 \alpha_{2} \alpha_{0}) \right) }}\nonumber\\
&
+{\frac { \left( d-2
 \right)  \left( d-3 \right)  \left( {r}^{2}+2\,{ b_1}\,{ \alpha_2}+
\sqrt { \left( b_{1}^{2}- b_{2}\right)4 \alpha_{2}^{2} + r^4(1-4 \alpha_{2} \alpha_{0})} \right)^{2}}{2{r}^{4}{
{ \alpha_2^2}}}}\nonumber\\
&
-2\,{\frac {R[\gamma] \left( {r}^{2}+2\,{ b_1}\,{ \alpha_2}+
\sqrt { \left( b_{1}^{2}- b_{2}\right)4 \alpha_{2}^{2} + r^4(1-4 \alpha_{2} \alpha_{0})} \right) }{{ \alpha_2}\,{r}^{4}}}+{\frac {K[\gamma]}{{r}^{
4}}}
\end{align}
which clearly has singularities at $r=0$ and at
\begin{equation}\label{sing2}
    r_s=\left ( \frac{(b_2-b_{1}^{2}) 4 \alpha_{2}^{2}}{1- 4 \alpha_{2} \alpha_{0}}\right)^{\frac{1}{4}}\; .
\end{equation}
Imposing the condition \eqref{largercond},
it is straightforward to show that the inner horizon $r_{-}(0,0) > r_s$ 
provided \eqref{cond2} holds; otherwise $r_{+}(0,0) > r_s$ 
if either of \eqref{cond1} or
\eqref{cond3} hold.
 If $b_2 < b_1^2$ the singularity in \eqref{sing2} is absent, but the $r=0$ singularity  in general remains.
 If $b_2 = b_1^2$ the spacetime is of constant curvature and all singularities are absent.

% \brh{Shouldn't it be when $b_{2}=b_{1}^{2}$ only $r=0$ is a singularity.} 

%\rbm{Note that $r_h^2$ is also real if $\alpha<0$ (de Sitter space) and $b_1 >0$.  This means that we have massless black holes in dS in Lovelock gravity!}

\subsection{Equation of State}

The equation of state   \eqref{eos} for $K=2$  is
\begin{equation}\label{eosK2}
P=\frac{(d-2) T}{4 r_{+}}-\frac{(d-2)(d-3) b_{1}}{16 \pi r_{+}^{2}}+\frac{(d-2) \alpha_{2} b_{1} T}{2 r_{+}^{3}}-\frac{(d-2)(d-5) \alpha_{2} b_{2}}{16 \pi r_{+}^{4}}+\frac{Q^{2}}{2 r_{+}^{2(d-2)}}.
\end{equation}
and becomes
\begin{equation}\label{eosdim}
p=\frac{t}{v}-\frac{(d-2)(d-3) b_{1}}{4 \pi v^{2}}+\frac{2 b_{1} t}{v^{3}}-\frac{(d-2)(d-5) b_{2}}{4 \pi v^{4}}+\frac{q^{2}}{v^{2}(d-2)}.
\end{equation}
upon introducing the   dimensionless variables $(v,t,m,q,p)$
\cite{Frassino_2014}
\begin{equation}\label{dimv}
r_{+}=v \alpha_{2}^{\frac{1}{2}}, \quad T=\frac{t \alpha_{2}^{-\frac{1}{2}}}{d-2}, \quad m=\frac{16 \pi M}{(d-2) \Sigma_{d-2}}{\alpha_{2}^{\frac{d-3}{2}}}, \quad Q=\frac{q}{\sqrt{2}} \alpha_{2}^{\frac{d-3}{2}}, \quad P=\frac{p}{4 \alpha_{2}}
\end{equation}
and recalling \eqref{maxP} it becomes
\begin{equation} \label{maxp}
p \leq p_{\max }=\frac{(d-1)(d-2)}{16 \pi }
\end{equation}.

Critical points are obtained by solving 
\begin{equation}
    \frac{\partial p}{\partial v}=0, \quad \frac{\partial^{2} p}{\partial v^{2}}=0 \label{criticalpointsequation}
\end{equation}
where the first equation determines critical temperature and the latter yields the critical volume. We obtain
\begin{equation}\label{critt}
t_{c}=\frac{\left(-4 v_{c}^{8-2 d} q^{2} \pi+b_{1}(d-3) v_{c}^{2}+2 b_{2}(d-5)\right)(d-2)}{2 \pi\left(v_{c}^{2}+6 b_{1}\right) v_{c}},
\end{equation}
and
\begin{equation}
\begin{split}\label{critv}
(48 v_{c}^{8-2 d} \pi d b_{1}+ & 8 v_{c}^{10-2 d} \pi d-168 v_{c}^{8-2 d} \pi b_{1}-20 v_{c}^{10-2 d} \pi) q^{2}+\\ &\left(-d b_{1}+3 b_{1}\right) v_{c}^{4}+\left(6 d b_{1}^{2}-6 b_{2} d-18 b_{1}^{2}+30 b_{2}\right) v_{c}^{2}-12 d b_{1} b_{2}+60 b_{1} b_{2}=0
\end{split}
\end{equation}
The  Gibbs free energy 
can also be written in dimensionless form using
\eqref{dimv} as 
\begin{equation}
    g=\frac{1}{\Sigma_{d-2}^{(\kappa)}} \alpha_{2}^{\frac{3-d}{2}} G
\end{equation}
yielding
\begin{equation}
\begin{split}
    g&=\frac{(d-2)\left(\frac{v^{d-2}}{d-2}+\frac{2 b_{1} v^{d-4}}{d-4}\right)\left(\frac{4 \pi p v^{2}}{d-2}+(d-3) b_{1}+\frac{(d-5) b_{2}}{v^{2}}\right)}{16 \pi v\left(1+\frac{2 b_{1}}{v^{2}}\right)} \\
    & +\frac{(d-2)\left(\frac{4 \pi p v^{d-1}}{(d-1)(d-2)}+b_{1} v^{d-3}+b_{2} v^{d-5}\right)}{16 \pi} \\
    & +\frac{q^{2}\left((2 d-5)(d-4) v^{2}+2 b_{1}(d-2)(2 d-7)\right)}{4(d-4)(d-2)(d-3)\left(v^{2}+2 b_{1}\right) v^{d-3}}
\end{split}
\end{equation}
from \eqref{gibbs0}.

The positive entropy condition \eqref{posentropy} for $K=2$ is 
\begin{equation}
   \frac{r_{+}^{d-2}}{d-2}+\frac{2 \alpha_{2} b_{1} r_{+}^{d-4}}{d-4} \geq 0 \Rightarrow
     \frac{v^{d-2}}{d-2}+\frac{2 b_{1} v^{d-4}}{d-4} \geq 0
\end{equation}
which is always satisfied for $b_1 > 0$. For
$b_1 < 0$
\begin{equation}
    v \geq \sqrt{2\frac{|b_{1}|(d-2)}{d-4}}
\end{equation}
implying that the size of the black hole must
be sufficiently large for it to have positive entropy.

Finally, we note that
the vacuum horizon equation \eqref{vachor} becomes
\begin{equation}
    v_{\pm}^{2}(0,0)=\frac{(d-1)(d-2)}{8 \pi p} \left( -b_{1} \pm \sqrt{b_{1}^{2} - \frac{16  \pi p b_{2}}{(d-1)(d-2)}} \right) 
\end{equation}
using the dimensionless variables \eqref{dimv}. 

\subsection{Five Dimensions}

 In light of our discussion in section 2.1, in five dimensions we must obey certain conditions in order to have compatibility with the constant curvature case. The conditions are $ b_{2}=b_{1}^{2}$, where 
$ b_{1} = -1,0,1$.  We shall consider only $b_1=\pm 1$ in what follows.  
 
 With this, we no longer have the possibility of a singularity outside of the origin in  vacuum spacetime (which we can  see in \eqref{sing2} with the $b_{2}=b_{1}^{2}$ condition). Our $f(r)$ solution for vaccuum is given by
\begin{equation}
    f_{\pm}(0,0)=\frac{r^{2}+2 b_{1} \alpha_{2} \pm \sqrt{r^{4}\left(1-4 \alpha_{2} \alpha_{0}\right)}}{2 \alpha_{2}}
\end{equation}
yielding
\begin{equation}
    r_{\pm}^{2}(0,0)=\frac{\left  (\pm \sqrt{1-4 \alpha_{2} \alpha_{0}}-1\right) b_{1}}{2 \alpha_{0}}.
\end{equation}
for the  horizons. 
Note that since $ 1 > 1-4 \alpha_{0} \alpha_{2}>0$, there will be two horizons provided $b_{1}<0$. For $1- 4 \alpha_{0} \alpha_{2}=0$, there are two coincident horizons
\begin{equation}
    r_{+}^{2}=r_{-}^{2}=\frac{- b_{1}}{2 \alpha_{0}}
\end{equation} 
 which corresponds to be being at maximum AdS pressure, where
$b_{1}<0$ or else there will be no horizons.
As $\alpha_{0} \rightarrow 0$, the only horizon is
\begin{equation}
    \lim_{\alpha_{0}\to\ 0} r^{2}_{+}(0,0)=- b_{1} \alpha_{2}
\end{equation}
for  $b_{1}<0$, or else no horizons are present. While the former horizon corresponds to a maximum AdS pressure the later corresponds to zero AdS pressure. Which shows that as we turn off the cosmological constant we will still maintain a vacuum singularity as long as we properly choose our topological term $b_{1}$.

For nonzero $M$ and $Q$ we obtain 
\eqref{elbhsol}
\begin{equation}\label{elbhsol5d}
f =  \frac{r^{2}+2 b_{1} \alpha_{2} \pm \sqrt{ r^4(1-4 \alpha_{2} \alpha_{0})
+ {8\textsf{m}\alpha_{2}} 
-\frac{4 \textsf{q}^2 \alpha_{2}}{r^{2}}
}}{2 \alpha_{2}}
\end{equation}
from \eqref{elbhsol}.  For $Q = 0$ there is a bound on the mass
\begin{equation}
 \textsf{m} \geq -\frac{b_{1}^{2}\left(1-4 \alpha_{0} \alpha_{2}\right)}{8 \alpha_{0}}
\end{equation} 
which provides a lower (negative) bound for the mass, below which uncharged black hole solutions do not exist.

\subsubsection{Uncharged ELBH Thermodynamics}

The five dimensional uncharged equation of state is given by
\begin{equation}
    p=\frac{t}{v}-\frac{3 b_{1}}{2 \pi v^{2}}+\frac{2 t b_{1}}{v^{3}}
\end{equation}
where 
\begin{equation}
 p_{max}=\frac{3}{4 \pi} \approx 0.2387324146 \label{maxp5d}
\end{equation}
is maximum dimensionless pressure.
The Gibbs free energy is
\begin{equation}
    g=-\frac{3\left(\frac{1}{3} v^{3}+2 b_{1} v\right)\left(\frac{4 \pi p v^{2}}{3}+2 b_{1}\right)}{16 \pi v\left(1+\frac{2 b_{1}}{v^{2}}\right)}+\frac{3\left(\frac{1}{3} \pi p v^{4}+b_{1} v^{2}+b_{2}\right)}{16 \pi}.
\end{equation}
The  critical temperature and volume are
obtained from \eqref{critt} and \eqref{critv}
\begin{equation}
    t_{c}=\frac{3 b_{1} v_{c}}{\pi\left(v_{c}^{2}+6 b_{1}\right)}, \quad   v_{c}^{2}-6 b_{1}=0.
\end{equation}
whose solutions are
\begin{equation}
 v_{c}=\sqrt{6 b_{1}}, \quad t_{c}=\frac{\sqrt{6 b_{1}}}{4 \pi}, \quad p_{c}=\frac{1}{12 \pi} = \frac{p_{max}}{9}.
\end{equation}
We  see that we must have positive values of $b_{1}$ in order to have real critical points;
this in turn ensures  the positive entropy condition holds for all black hole sizes.

\begin{figure}[H]
    \centering
    \begin{subfigure}{0.29\textwidth}
    \centering
    \includegraphics[width=\textwidth]{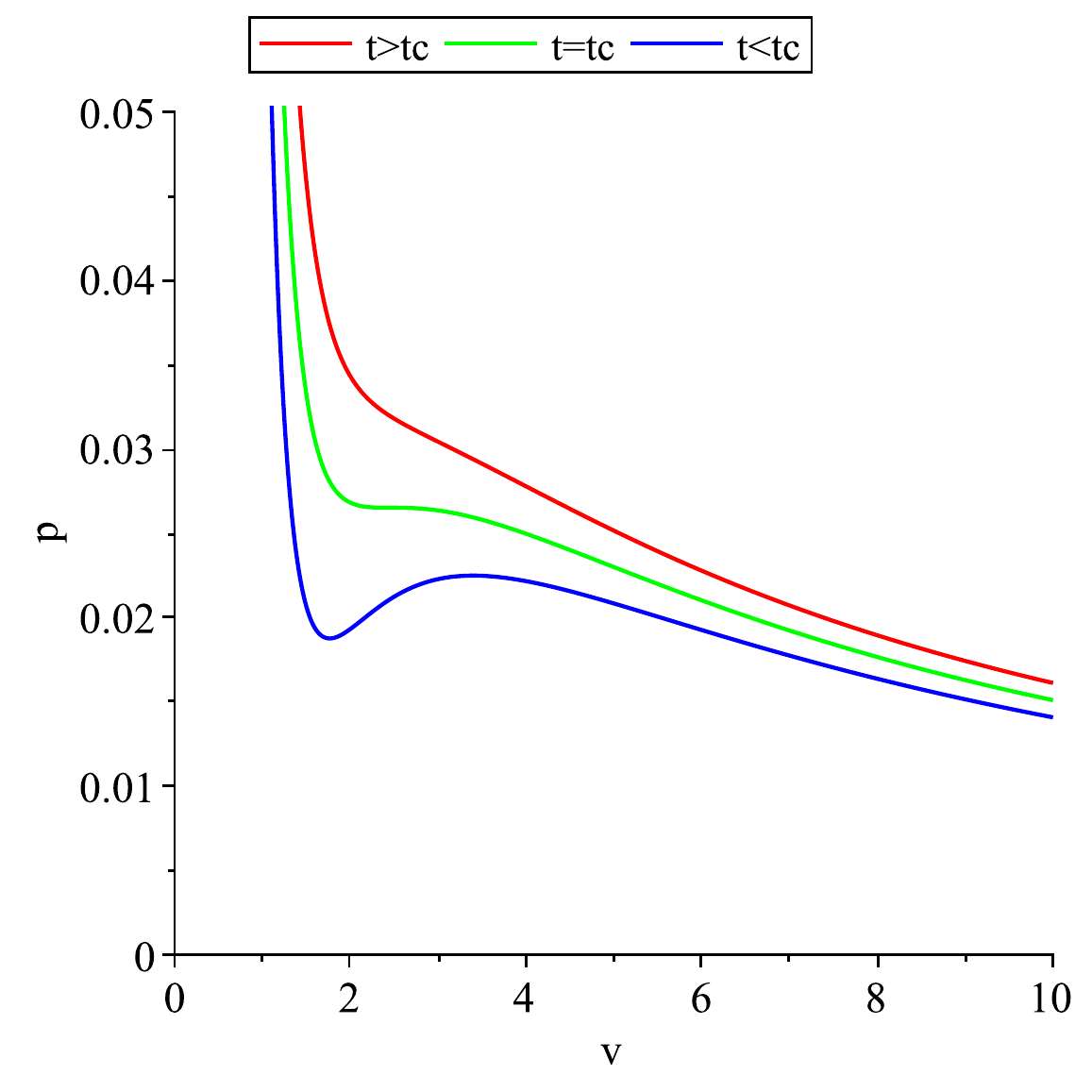}
    \end{subfigure}
    \begin{subfigure}{0.29\textwidth}
    \centering
    \includegraphics[width=\textwidth]{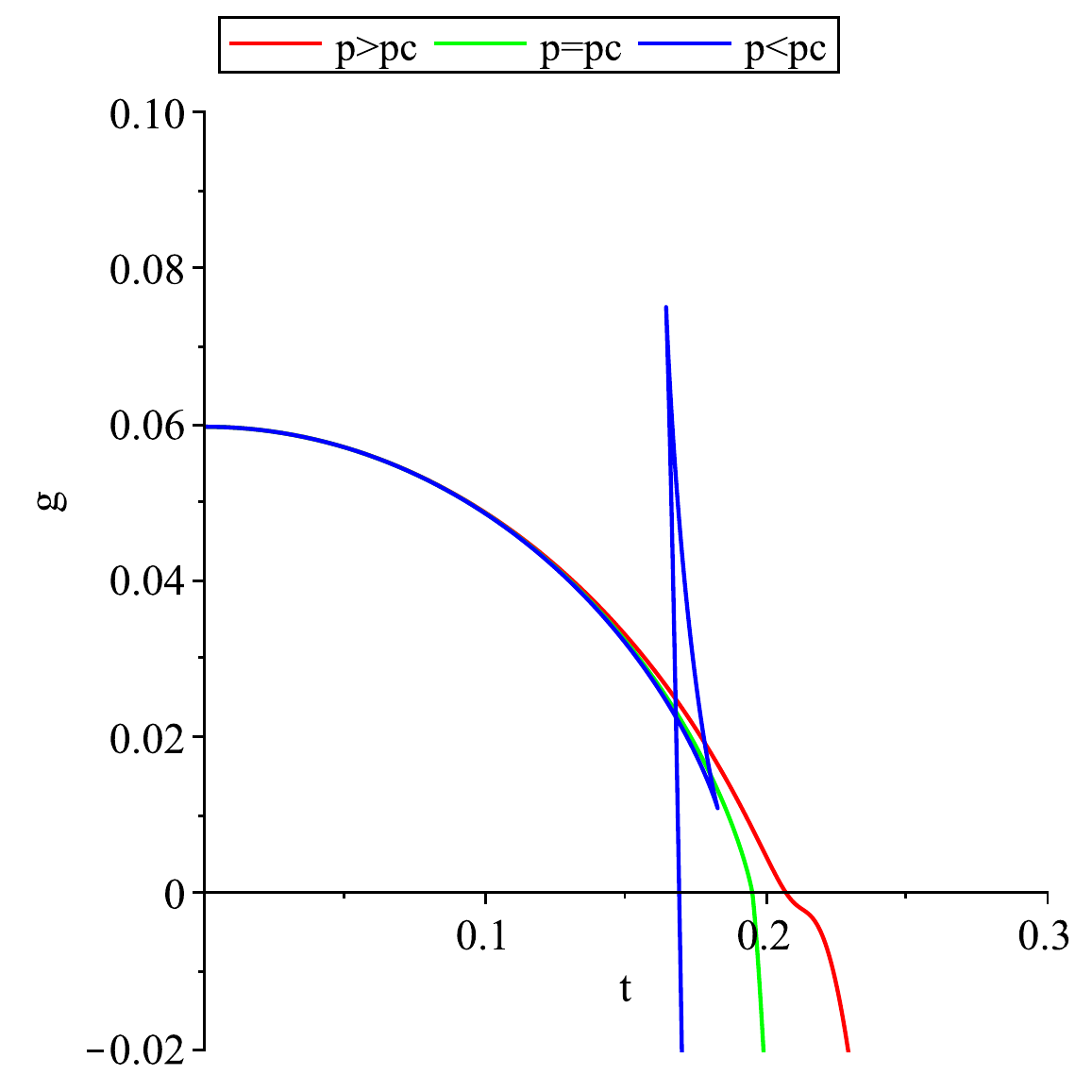}
    \end{subfigure}
    \begin{subfigure}{0.29\textwidth}
    \centering
    \includegraphics[width=\textwidth]{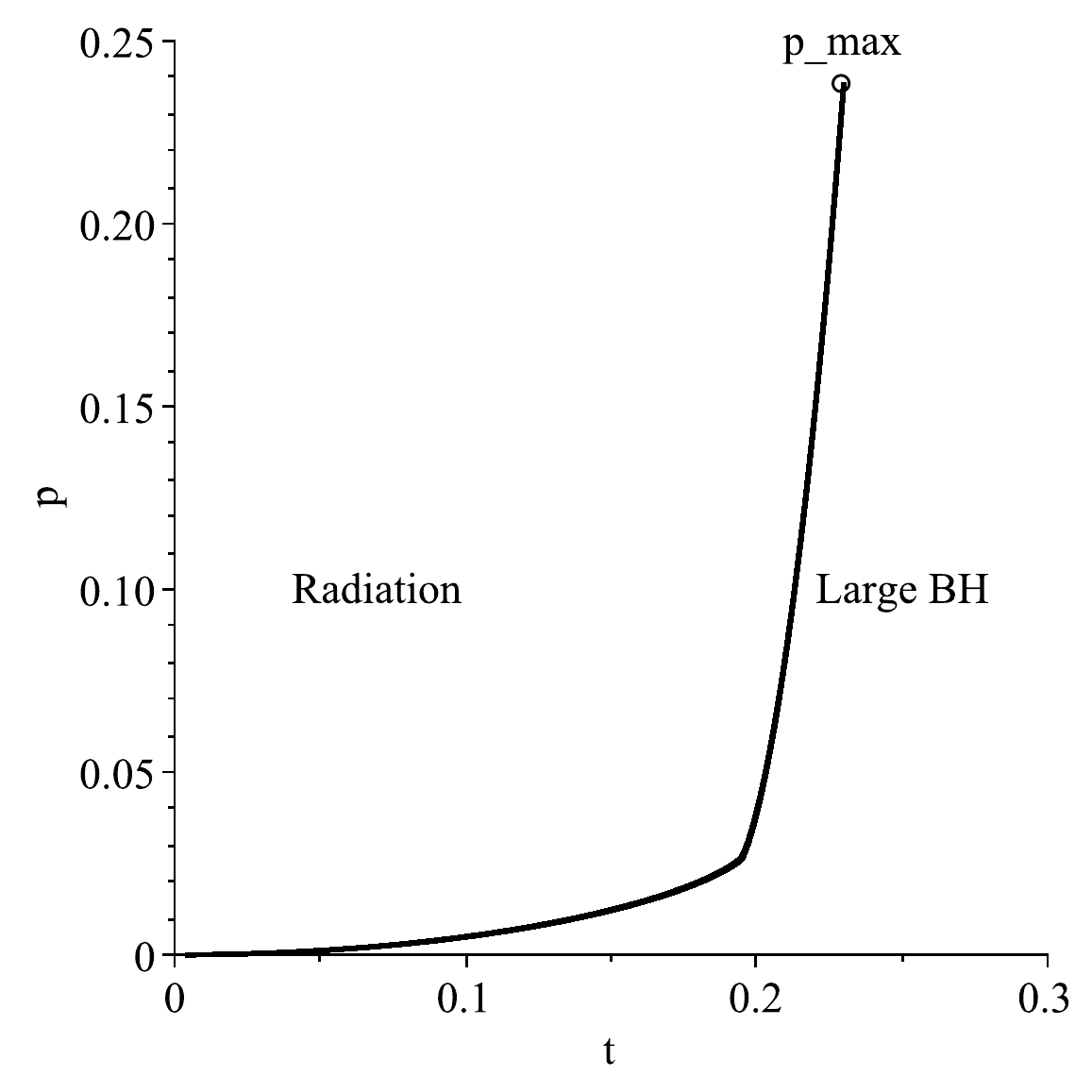}
    \end{subfigure}
    \caption{\textbf{Phase Behaviour for $d=5$,  $q=0$, $b_{1}=b_{2}=1$ Black Holes.} \textit{Left:} $p-v$ diagram with constant temperature slices of the \textit{unstable BH} showing the oscillation for $t<t_{c}$. \textit{Center}: $g-t$ diagram with constant pressure slices around $p_{c}$ again, of \textit{unstable BHs} showing swallowtail structure with intersection between large and small black holes. \textit{Right:} Phase diagram of the Black Hole / Radiation showing the termination at the maximum pressure.}
    \label{5DUncharged1}
\end{figure}

\begin{figure}[H]
    \centering
    \includegraphics[width=0.5\textwidth]{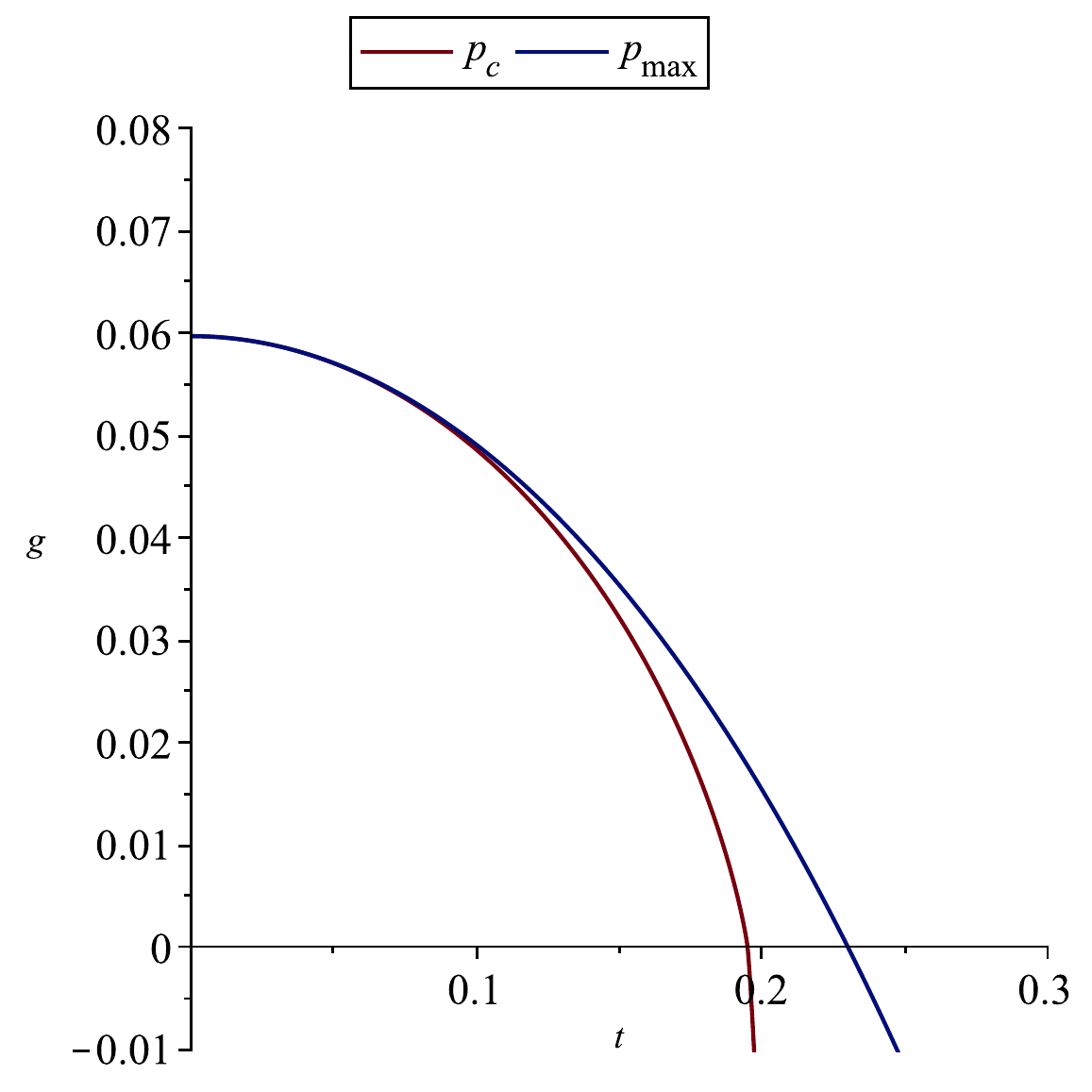}
    \caption{$g-t$ plots for two constant pressure slices: $p_{crit}$ of the unstable black hole and $p_{max}$ of the spacetime.}
    \label{pc/pmax-gt}
\end{figure}

With the condition that $b_{1}^{2}=b_{2}$ we can first begin with the constant curvature case, $b_{1}=b_{2}=1$. We see the standard Van der Waals behaviour\cite{Frassino_2014}  shown in Figure~\ref{5DUncharged1}. Seen in the center image there is an intersection between large and small black hole branches, indicating  a large/small first order phase transition.  However these black holes are unstable since their free energy is greater than that of $g=0$ which corresponds to AdS radiation. Instead, when the large black hole branch crosses the $g=0$ axis it will undergo a Hawking/Page transition into thermal AdS radiation \cite{Hawking:1982dh}. The phase diagram for this first order transition is displayed in the right image of Figure~\ref{5DUncharged1}.
 The apparent sharp corner occurs at the critical point $p=p_c$ for the unstable black hole branch. It is actually smooth, but corresponds to a very steep rise in pressure as a function of temperature for 
$p_c < p < p_{max}$ from the equation of state, as shown in Figure~\ref{pc/pmax-gt} .

\begin{figure}[H]
    \centering
    \begin{subfigure}{0.4\textwidth}
    \centering
    \includegraphics[width=\textwidth]{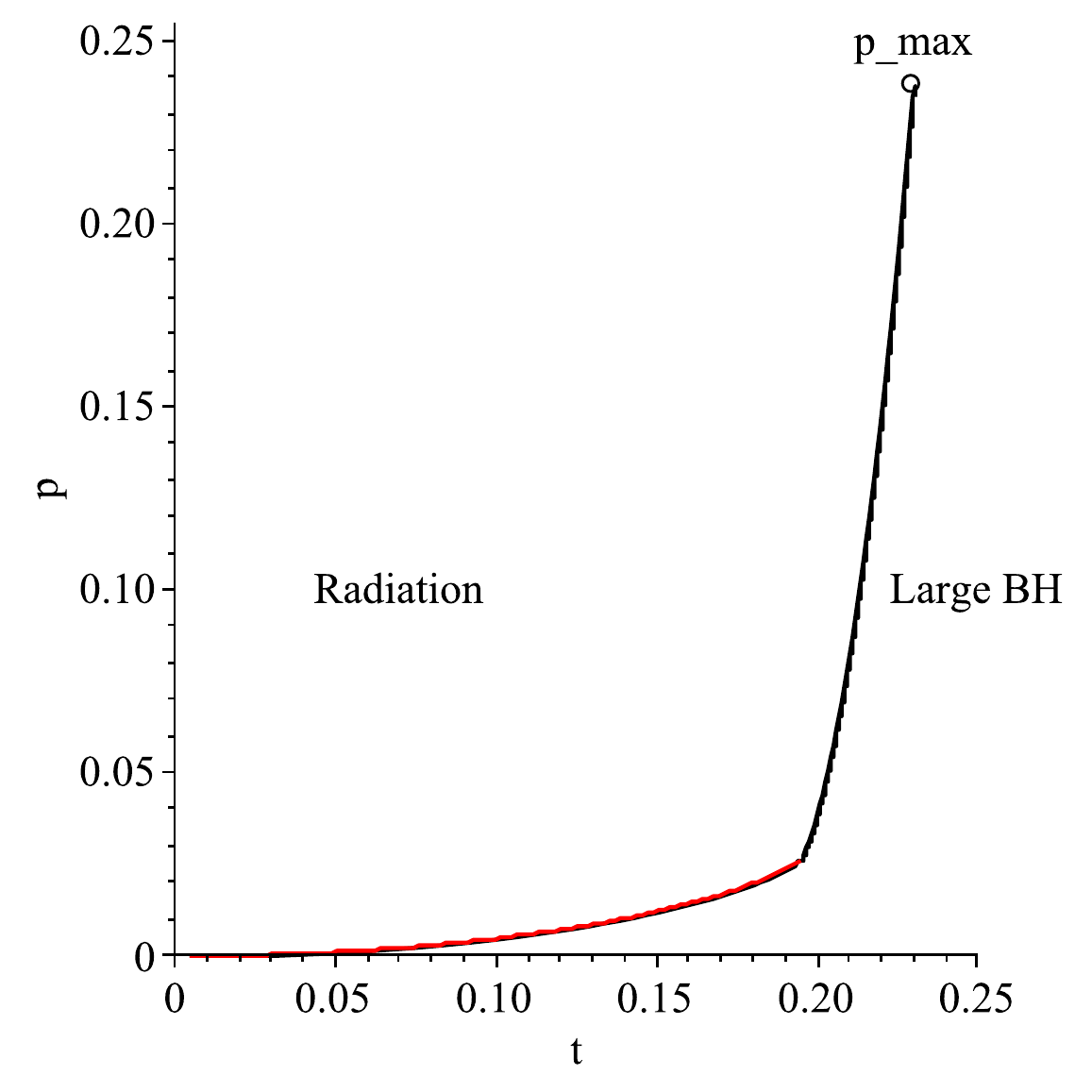}
     \end{subfigure}
     \centering
    \begin{subfigure}{0.4\textwidth}
    \centering
     \includegraphics[width=\textwidth]{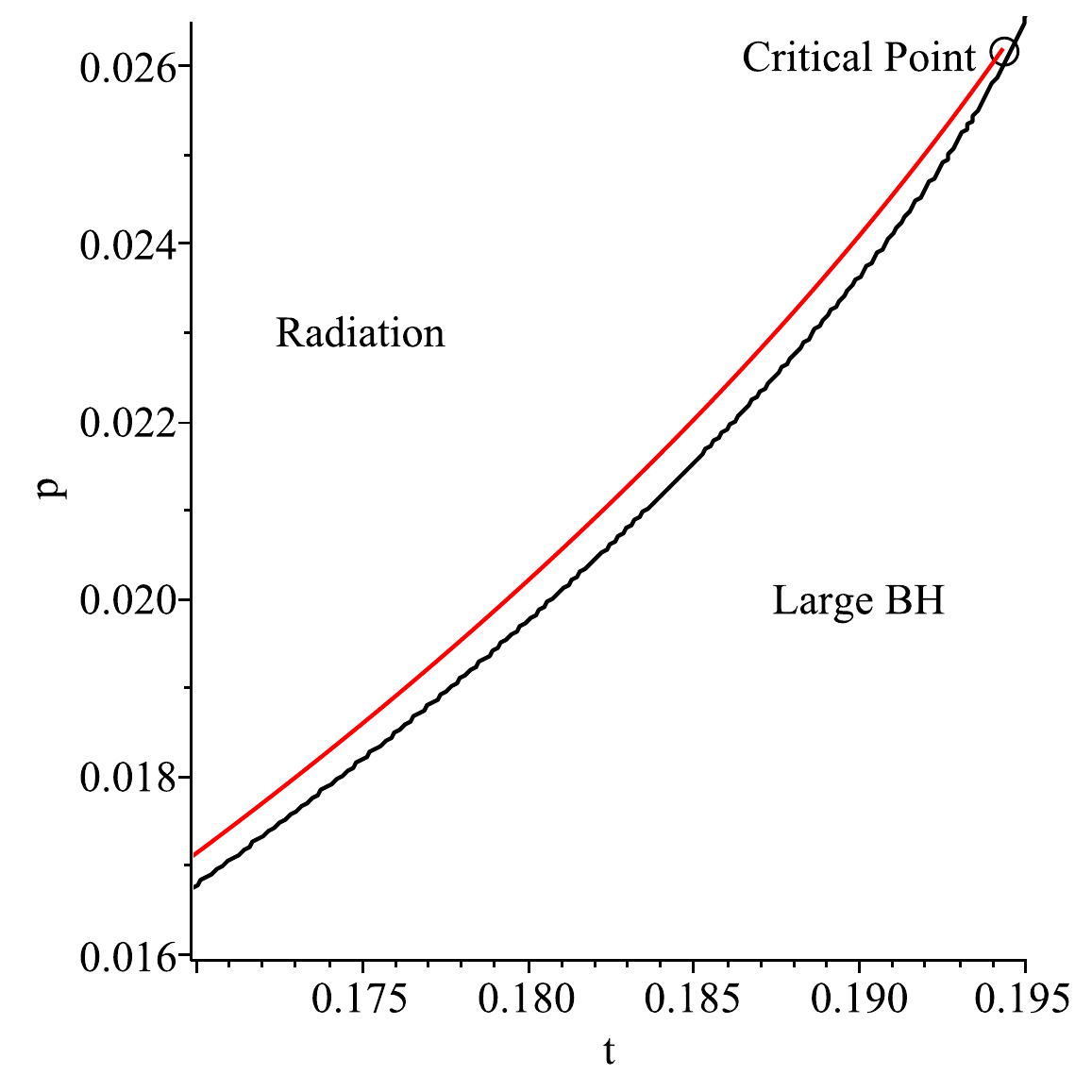}
      \end{subfigure}
    \caption{Left: Phase diagram showing the first-order coexistence line of the unstable large/small transition (red) and the coexistence line of the Hawking-Page transition(black). Right: Close up version of the left diagram. The unstable small black hole phase is between the red and black lines. }
    \label{fig:compare}
\end{figure}

 It is interesting to compare the phase behaviour of considered previously for uncharged $d=5$ Gauss-Bonnet black holes with the present case. As shown in Figure~\ref{fig:compare}, the coexistence line between the radiation/large black hole  phases is very close to that of the small/large black hole phases.  Approaching the diagram from the right, it is clear that as the temperature decreases, the large black hole will undergo a phase transition to radiation before that for a small black hole.

\subsubsection{Charged ELBH Thermodynamics}

Including charge, the equation of state and Gibbs free energy are now
\begin{equation}
    p=\frac{t}{v}-\frac{3 b_{1}}{2 \pi v^{2}}+\frac{2 t b_{1}}{v^{3}}+\frac{q^{2}}{v^{6}}
\end{equation}
\begin{equation}
 g=-\frac{3\left(\frac{1}{3} v^{3}+2 b_{1} v\right)\left(\frac{4 \pi p v^{2}}{3}+2 b_{1}\right)}{16 \pi v\left(1+\frac{2 b_{1}}{v^{2}}\right)}+\frac{3\left(\frac{1}{3} \pi p v^{4}+b_{1} v^{2}+b_{2}\right)}{16 \pi}+\frac{q^{2}\left(5 v^{2}+18 b_{1}\right)}{24\left(v^{2}+2 b_{1}\right) v^{2}}
\end{equation}
and the critical temperature equation and critical volume relations \eqref{critt} and \eqref{critv} become
\begin{equation}\label{tvcrit5dq}
    t_{c}=\frac{3\left(b_{1} v_{c}^{4}-2 \pi q^{2}\right)}{\pi v_{c}^{3}\left(v_{c}^{2}+6 b_{1}\right)}, \quad 3 v_{c}^{6} b_{1}-18 v_{c}^{4} b_{1}^{2}-\left(30 \pi v_{c}^{2}+108 \pi b_{1}\right) q^{2}=0.
\end{equation}
Since the latter is a cubic equation in $v_c^2$,
analytic solutions  are possible for arbitrary values of $b_{1}$ and $q$. All roots of the cubic will be positive and real only if its coefficients alternate in sign, which is not possible for any values of $b_1$ or $q$.  Hence there can be at most two admissible solutions for $v_c$ from \eqref{tvcrit5dq}.
%Of the 6 solutions, 4 can be eliminated immediately as they are complex for all choices of $b_{1}$ and $q$. 

\begin{figure}[H]
    \centering
    \begin{subfigure}{0.29\textwidth}
    \centering
    \includegraphics[width=\textwidth]{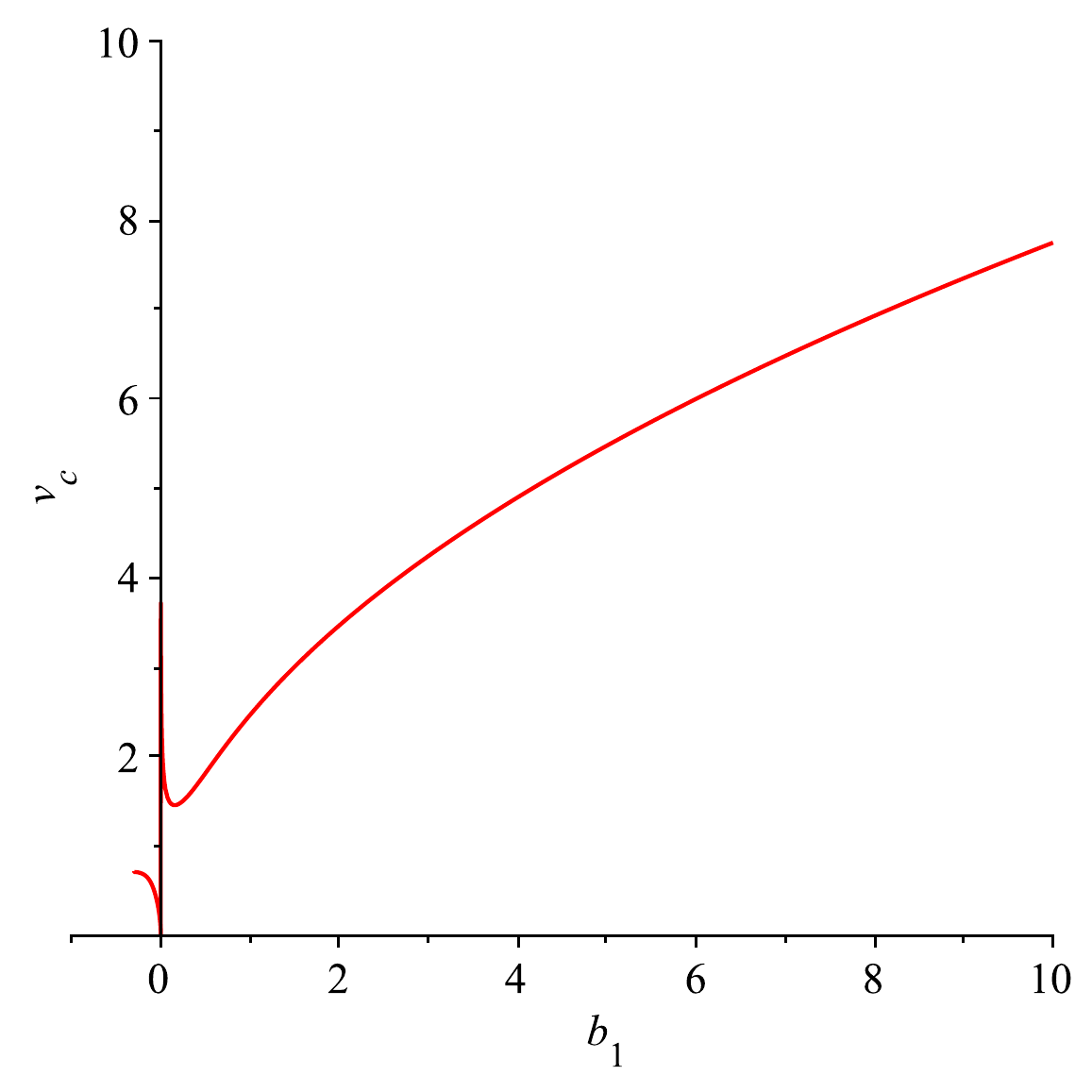}
    \caption{$q=0.1$}
    \end{subfigure}
    \begin{subfigure}{0.29\textwidth}
    \centering
    \includegraphics[width=\textwidth]{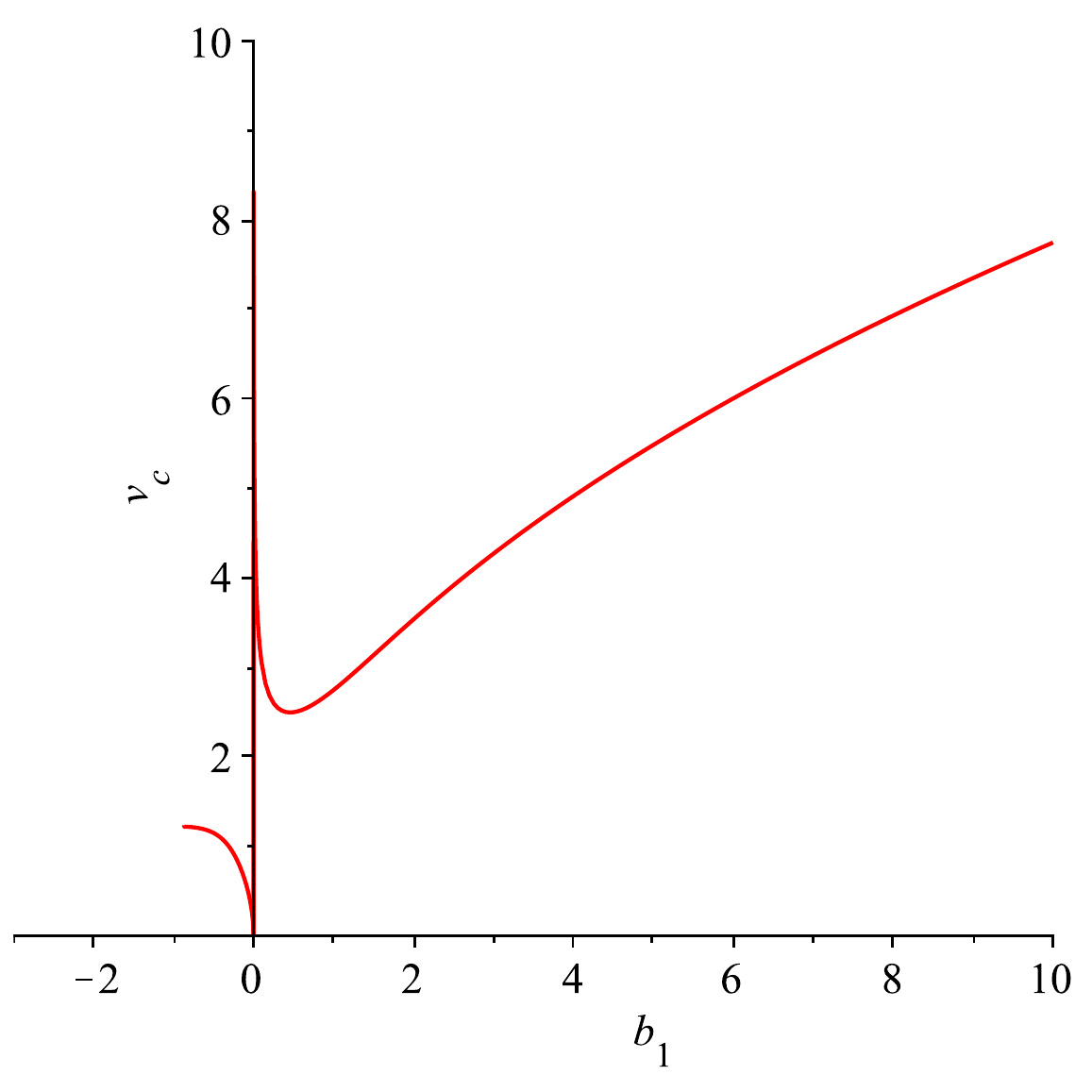}
    \caption{$q=0.5$}
    \end{subfigure}
    \begin{subfigure}{0.29\textwidth}
    \centering
    \includegraphics[width=\textwidth]{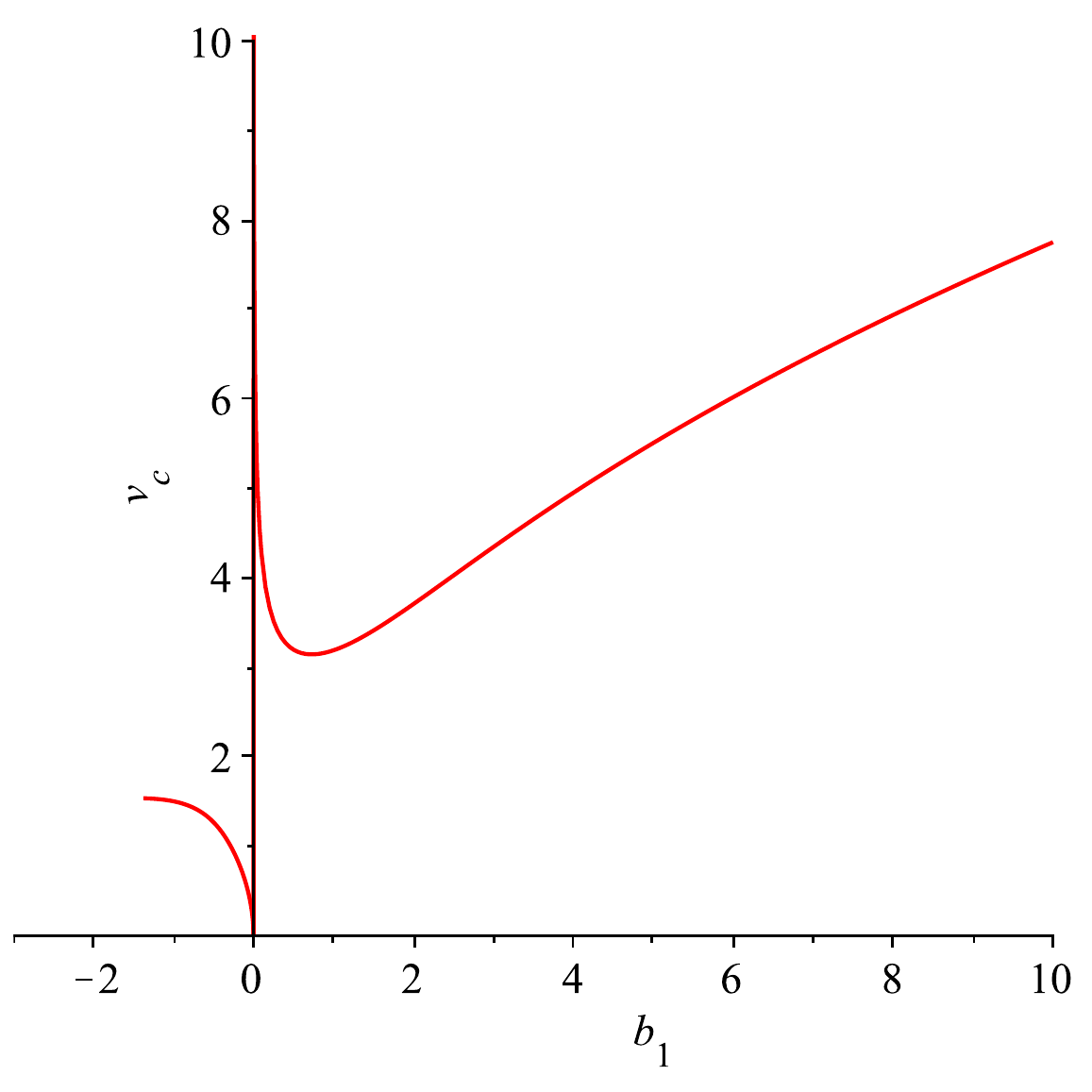}
    \caption{$q=1$}
    \end{subfigure}
    \caption{Critical volume solutions vs $b_{1}$ for varying values of $q$, the topological parameter $b_{2}$ does not play a role in these solutions.}
    \label{5dqvolplots}
\end{figure}

Plotting in Figure~\ref{5dqvolplots} the critical volume solutions for specific choices of $q$,  we  see that negative values of $b_{1}$ are permitted as well as positive ones,  unlike the uncharged case. For any given charge  there is a negative value of $b_{1}$
below which there are no longer any real solutions for $v_c$ from \eqref{tvcrit5dq}; for $q=1$ this is approximately $b_1 = -1.5$. As we decrease the charge, the magnitude of the most negative allowed value of $b_{1}$ also decreases, as expected since  only positive values of $b_{1}$ are permitted for  $q=0$.

For $b_1>0$, there is always a positive root to the cubic, so all positive values of $b_1$ yield a positive critical volume. 

\begin{figure}[H]
   \centering
    \begin{subfigure}{0.29\textwidth}
    \centering
    \includegraphics[width=\textwidth]{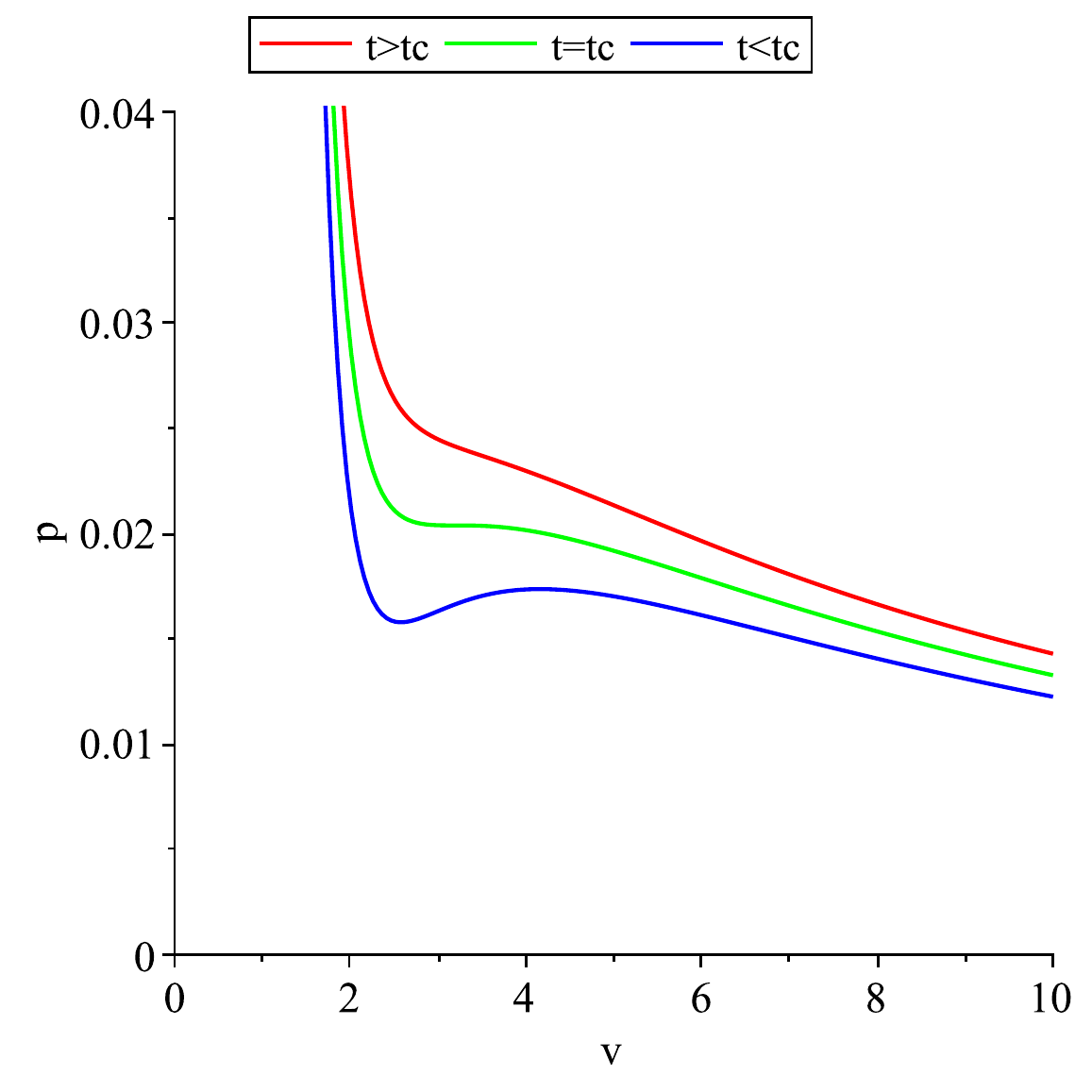}
    \end{subfigure}
    \begin{subfigure}{0.3\textwidth}
    \centering
    \includegraphics[width=\textwidth]{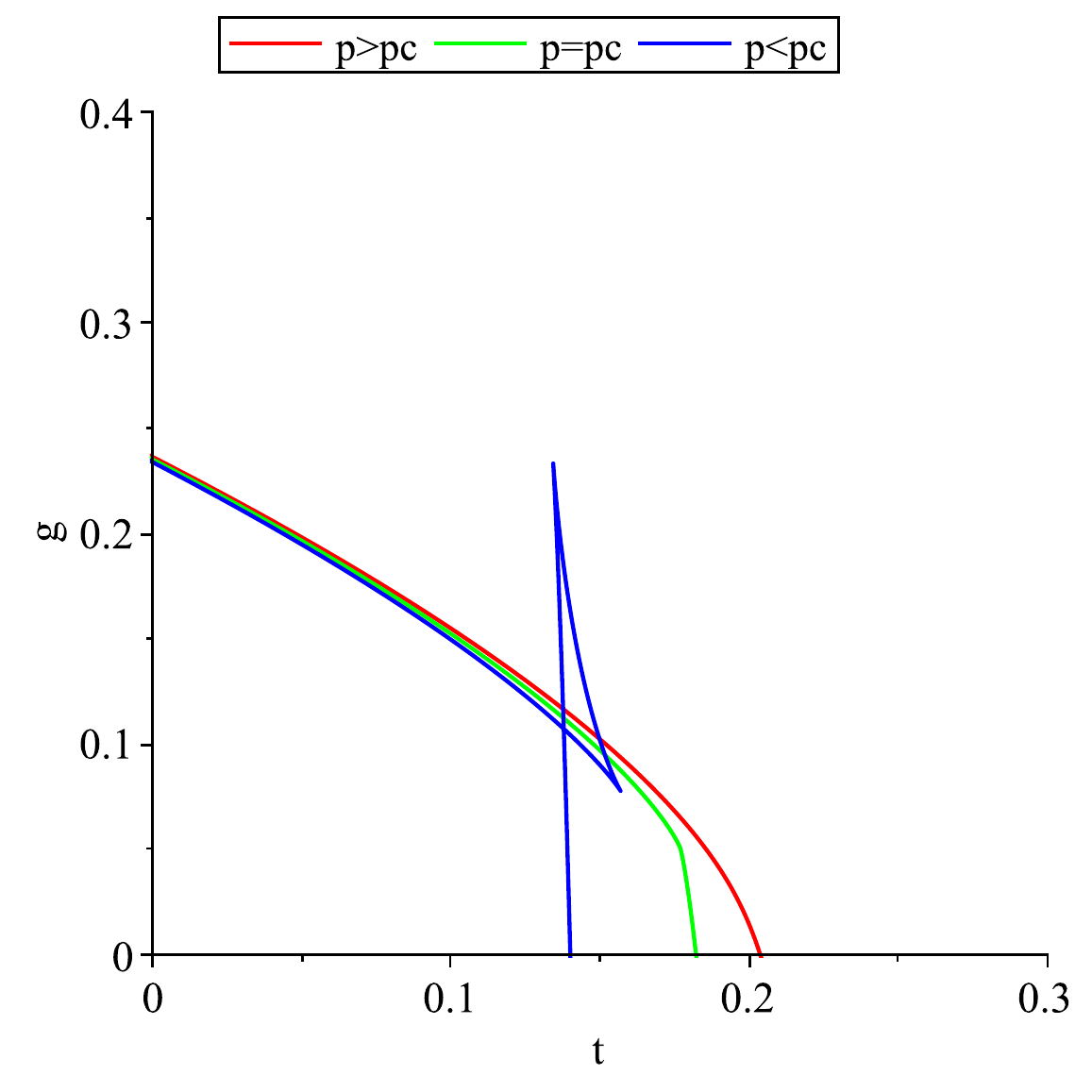}
    \end{subfigure}
    \begin{subfigure}{0.29\textwidth}
    \centering
    \includegraphics[width=\textwidth]{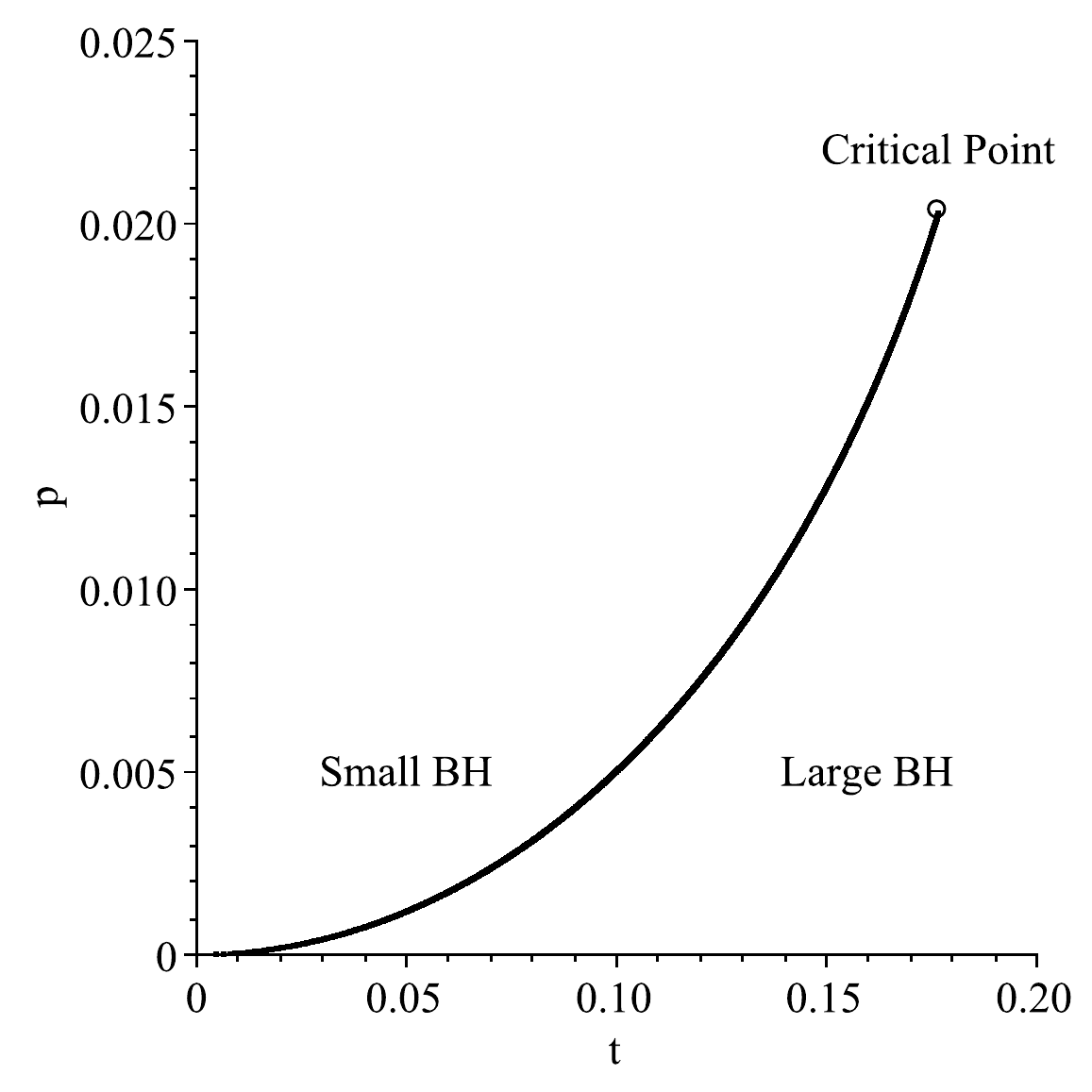}
    \end{subfigure}
    \caption{\textbf{Phase Behaviour for $d=5$, $q=1$\,,\,$b_{1}=b_{2}=1$ black holes.} \textit{Left:} $p-v$ diagram with constant temperature values mimicking the uncharged case with Van der Waals oscillation. \textit{Bottom:} $g-t$ diagram with constant pressure slices showing large/small branch intersection. \textit{Right:} phase diagram displaying first order transition terminating at the critical point.}
    \label{pvgt5dq(1)}
\end{figure}

We   first begin with $q=1$, $b_{1}=b_{2}=1$, illustrating the results in Figure~\ref{pvgt5dq(1)}. In this configuration they exhibit the standard Van der Waals behaviour of a large/small black hole phase transition \cite{Frassino_2014} . There is   no transition into radiation (thermal AdS)  due to conservation of charge. This behaviour is qualitatively the same for all possible values of $b_1>0$.

%Therefore the first order transition occurring at the branch intersections of the Gibbs free energy vs. temperature plot are indeed a Large/Small black hole transition which is displayed in the right figure. For all possible positive values of $b_{1}>b_{1}_{min}$ the large/small transition is the only possible option. Therefore further explorations of different values of $b_{1}$ is not warranted due to no new effects. 

\begin{figure}[H]
    \centering
    \begin{subfigure}{0.45\textwidth}
    \centering
    \includegraphics[width=\textwidth]{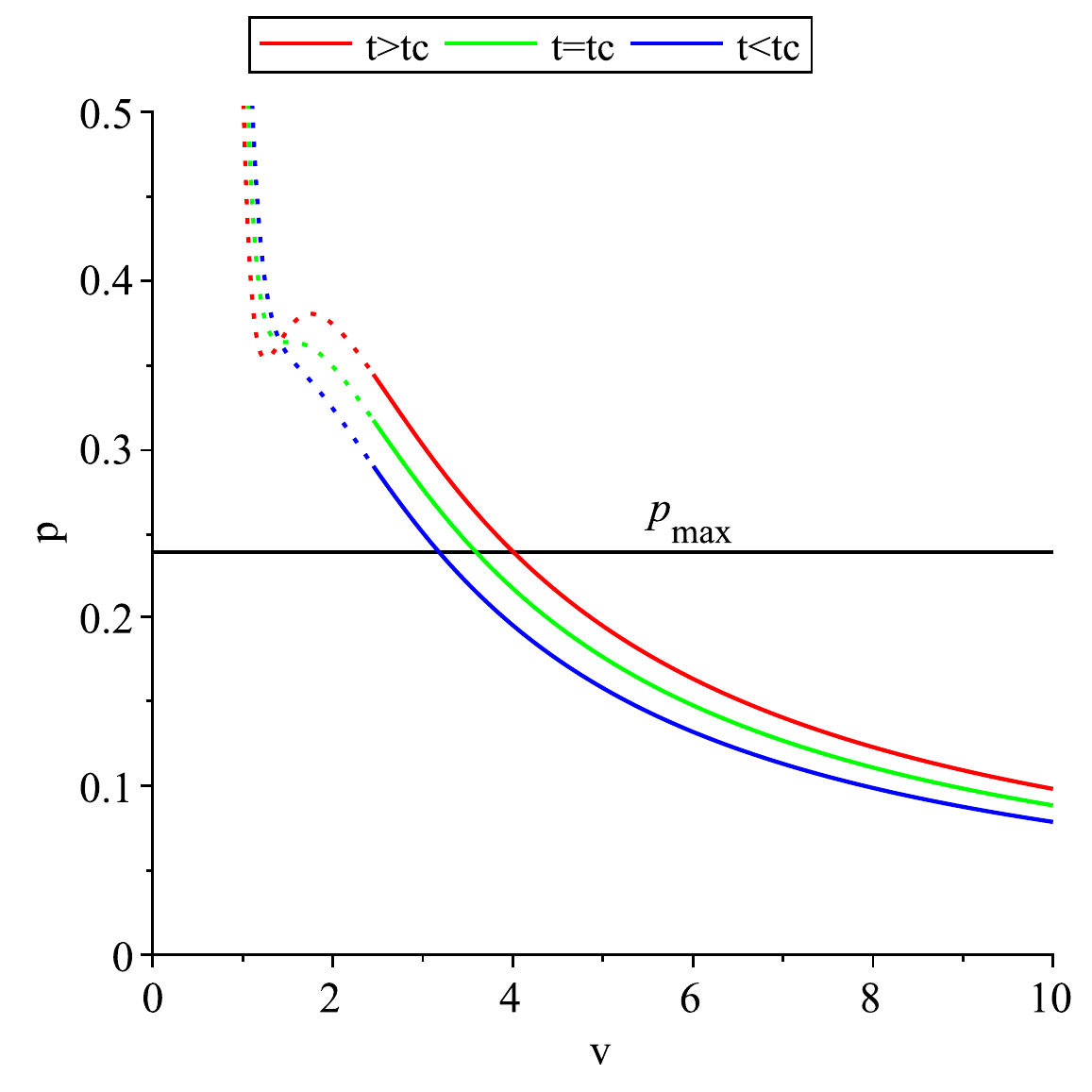}
    \end{subfigure}
    \begin{subfigure}{0.45\textwidth}
    \centering
    \includegraphics[width=\textwidth]{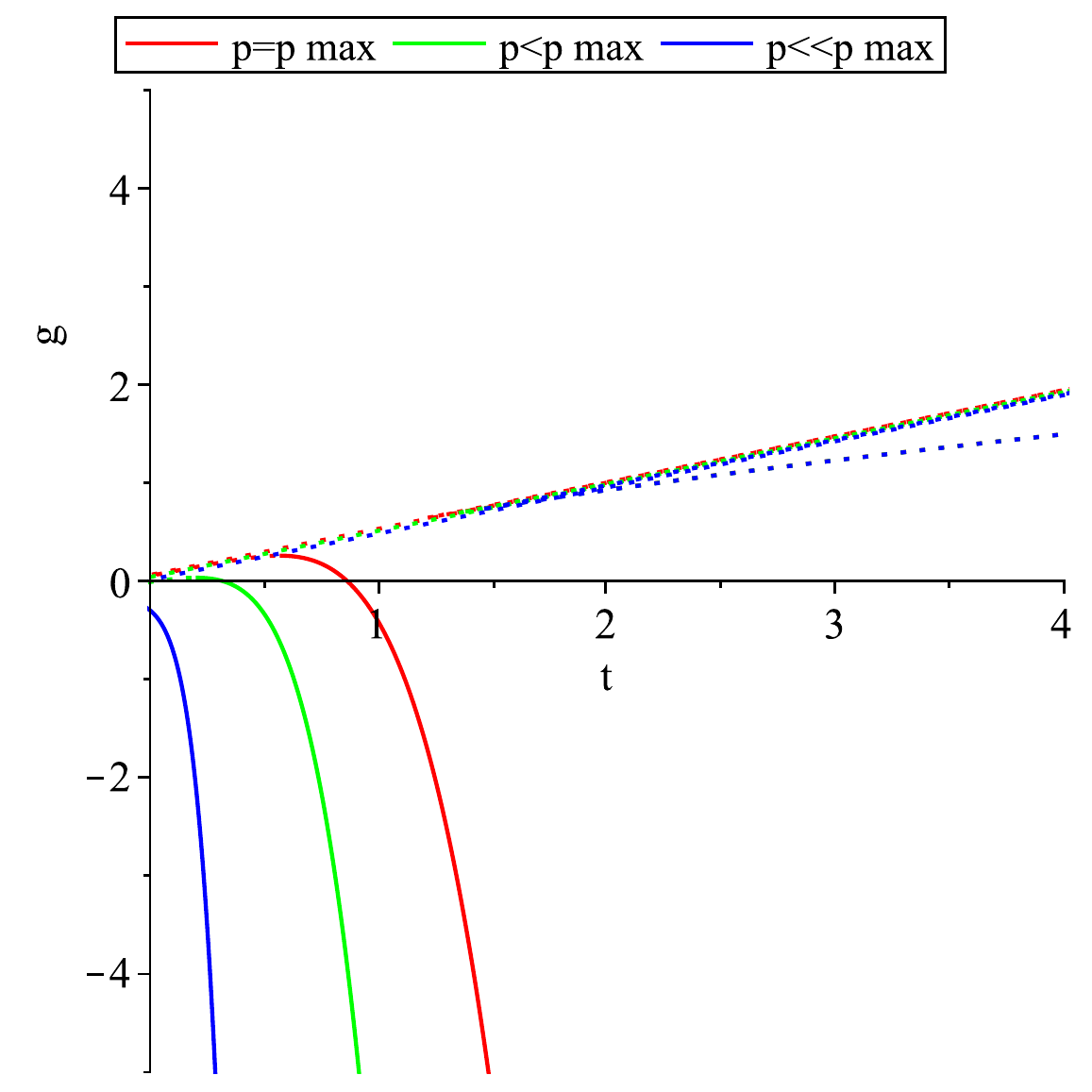}
    \end{subfigure}
    \caption{\textbf{Phase Behaviour for $d=5$, $q=1$, $b_{1}=-1,\: b_{2}=1$ black holes.} \textit{Left:} $p-v$ diagram for constant temperature slices. \textit{Right:} $g-t$ diagram for constant pressure slices beginning with maximum pressure and decreasing. Solid lines represent black holes with positive entropy while dotted lines correspond to negative entropy}
    \label{pvgt5db1-1}
\end{figure}

We illustrate the situation for $b_1=-1$ (and $b_2=1$) in Figure~\ref{pvgt5db1-1}. In this case there is no Van der Waals type behaviour for $p<p_{max}$, and no interesting phase behaviour.

%As the critical volume solutions allow, we will examine when we have $b_{1}<0$. Being in 5 dimensions we will look with $b_{1}=-1, \: b_{2}=1$. We can see from the critical volume figures above that we are allowed to have to $b_{1}=-1$ within a certain range of q where we will keep our critical volume entirely real. However, there are other solution of critical volume which have complex contributions of the order of $10^{-11}$ which you could still carry out analysis on. With that being said we will stay where there are no complex contributions in the critical volume. First off beginning with $q=1$. We see our results in Figure~\ref{pvgt5db1-1}. What we can see from these figures is that there is no phase transition behaviour occurring when we have our topological parameter $b_{1}=-1$.

\subsection{Six Dimensions}

In six dimensions we are  free to choose our own values of $b_{1}$ and $b_{2}$.  In this case a vacuum singularity could occur at 
\begin{equation}
    v=\left ( \frac{4 (b_2-b_{1}^{2})  }{1 - \frac{4 p \pi}{5} }\right)^{\frac{1}{4}}
\end{equation}
provided $b_{2}-b_{1}^2>0$,
since $p < p_{max} =\frac{5}{4\pi} = 0.3978873576$.
The  vacuum horizon equation \eqref{vachor} can be rewritten as
\begin{equation}
   v_{\pm}^{2}(0,0)= \frac{5}{2 \pi p} \left( -b_{1} \pm \sqrt{b_{1}^2 - \frac{4 \pi p b_{2}}{5}}\right)
\end{equation}
using the dimensionless variables \eqref{dimv}.  

Now we can re-write the vacuum horizon conditions 
into dimensionless form as follows:
\begin{align}
\textrm{(a)}&\quad  b_2<0 \quad\textrm{and}\quad b_1>0\quad
\Rightarrow   v_{+}(0,0) \;\textrm{is the only horizon} \\
& \nonumber
\\
\textrm{or, if}\quad   
 b_2 > 0 \nonumber\\ 
\textrm{(b)}&\quad  b_1^2 > \frac{4 \pi p b_{2}}{5} \quad \textrm{and}\quad 0 >  b_1 > 
- \sqrt{\frac{b_{2}}{2 \left( 1 - \frac{2 \pi p}{5}\right) }}
\quad
\Rightarrow  v_{\pm}(0,0) \;\textrm{are both horizons}
\label{cond2a}
\\ 
\textrm{or}
\nonumber\\ 
\textrm{(c)}&\quad  b_1^2 > \frac{4 \pi p b_{2}}{5} \quad \textrm{and}\quad  - \sqrt{\frac{b_{2}}{2 \left(1 - \frac{2 \pi p}{5} \right)}} > b_1  \quad \Rightarrow v_{+}(0,0) \;\textrm{is the only horizon}
\label{cond3a}
\end{align}

For nonzero $M$ and $Q$ we have 
\begin{equation}\label{elbhsol6d}
f =  \frac{r^{2}+2 b_{1} \alpha_{2} \pm \sqrt{\left(b_{1}^{2}-b_{2}\right) 4 \alpha_{2}^{2}+ r^4(1-4 \alpha_{2} \alpha_{0})
+ \frac{{8\textsf{m}\alpha_{2}}}{r}
-\frac{4 \textsf{q}^2 \alpha_{2}}{r^{4}}
}}{2 \alpha_{2}}
\end{equation}
from \eqref{elbhsol}.  For $Q = 0$ we obtain

\begin{equation}
    \textsf{m} \geq \textsf{m}_{\pm} \equiv - \frac{\sqrt{10}\sqrt{-\alpha_{0}\left(3 b_{1} \pm \sqrt{-20 \alpha_{0} \alpha_{2} b_{2}+9 b_{1}^{2}}\right)} \left ( - 3 b_{1}^{2} \mp \sqrt{9 b_{1}^{2} - 20 \alpha_{0} \alpha_{2} b_{2}}+ 20 \alpha_{0} \alpha_{2} b_{2}\right)}{500 \alpha_{0}^{2}}
\end{equation}
as a lower (negative) bound for the mass, below which uncharged black hole solutions do not exist, with
$\textsf{m}_+$ corresponding to $b_1>0$ and
$\textsf{m}_-$ to $b_1<0$.  
Note that $\textsf{m}_+ > \textsf{m}_{-}$ (solutions with
$b_1<0$ can have more negative mass) and that
$b_2 \alpha_2 \alpha_0 < 9 b_1^2/20$ for such solutions to exist. If $b_2\alpha_2 >0$ then
only $\textsf{m}_{-} < 0$.

\subsubsection{Uncharged ELBHs}

The equation of state and Gibbs free energy for $d=6$ and $q=0$ are  
\begin{equation}
p=\frac{t}{v}-\frac{3 b_{1}}{\pi v^{2}}+\frac{2 t b_{1}}{v^{3}}-\frac{b_{2}}{\pi v^{4}}
\end{equation}
\begin{equation}
    g= -\frac{\left(\frac{1}{4} v^{4}+b_{1} v^{2}\right)\left(\pi p v^{2}+3 b_{1}+\frac{b_{2}}{v^{2}}\right)}{4 \pi v\left(1+\frac{2 b_{1}}{v^{2}}\right)}+\frac{\frac{1}{5} \pi p v^{5}+b_{1} v^{3}+b_{2} v}{4 \pi}
\end{equation}
The critical temperature relation \eqref{critt} is now
\begin{equation}
    t_{c}=\frac{2\left(3 b_{1} v_{c}^{2}+2 b_{2}\right)}{\pi v_{c}\left(v_{c}^{2}+6 b_{1}\right)}
\end{equation}
and the critical volume relation is \eqref{critv} is
\begin{equation}
    6 v_{c}^{4} b_{1}+\left(-36 b_{1}^{2}+12 b_{2}\right) v_{c}^{2}+24 b_{1} b_{2}=0 \label{critv6d}
\end{equation}
whose solutions  
are \begin{equation}
   v^2_{c\pm} =   \frac{ {\left(3 b_{1}^{2}\pm\sqrt{(9 b_{1}^{2}-b_{2})( b_{1}^2-b_{2})}-b_{2}\right)}}{b_{1}}
\end{equation}

If $b_{1}>0$ then we must have
$b_2 \leq b_1^2$ in order that
$v_{c}$  be real and positive. 
If $b_2 >0$, then both 
$v_{c\pm}$ are valid critical solutions.  Conversely, if $b_1<0$ then we must have either
$b_2<0$ or $b_2 > 9 b^2_1$; if the latter holds  both 
$v_{c\pm}$ are valid critical solutions.

\begin{figure}[H]
    \centering
    \begin{subfigure}{0.29\textwidth}
    \centering
    \includegraphics[width=\textwidth]{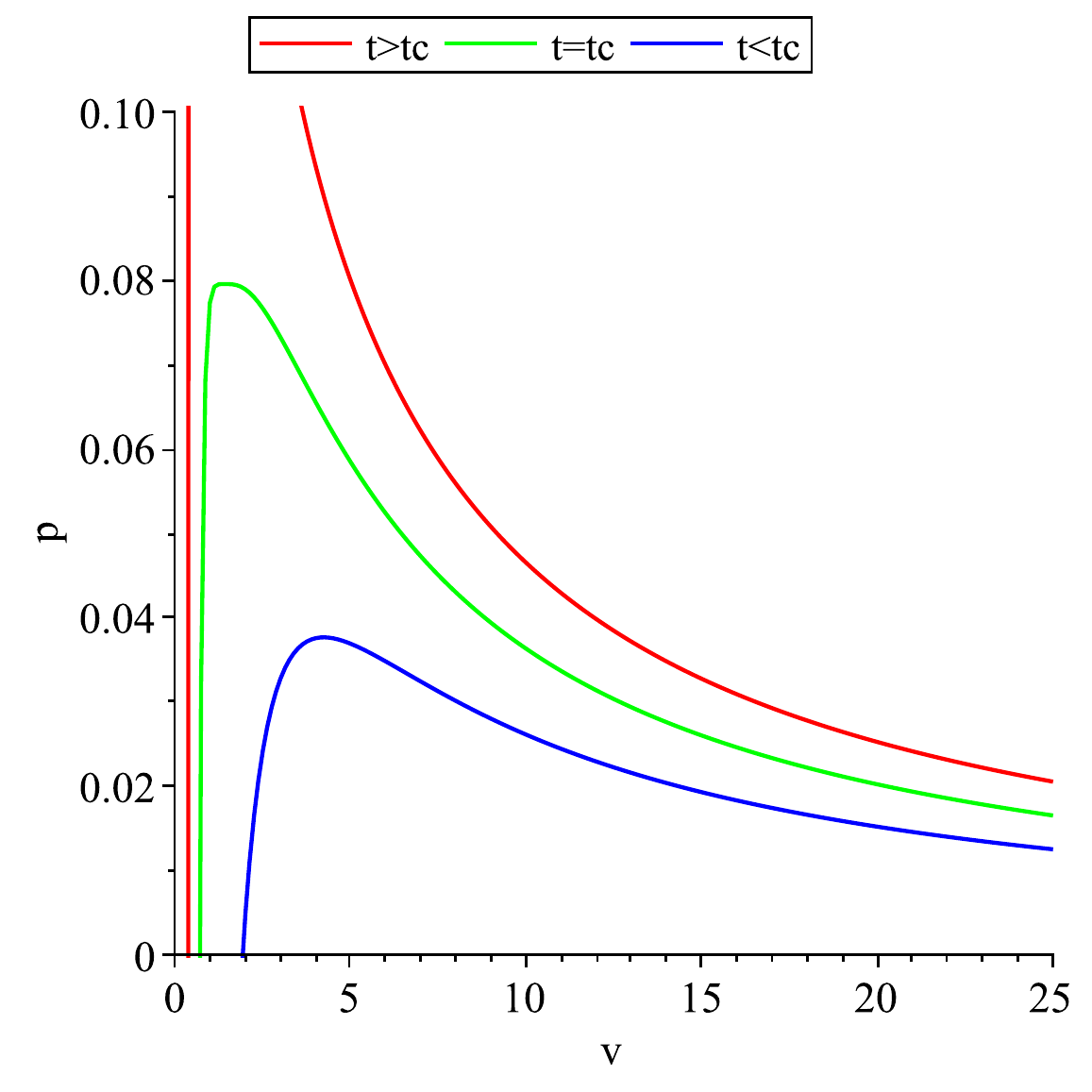}
    \end{subfigure}
    \begin{subfigure}{0.29\textwidth}
    \centering
    \includegraphics[width=\textwidth]{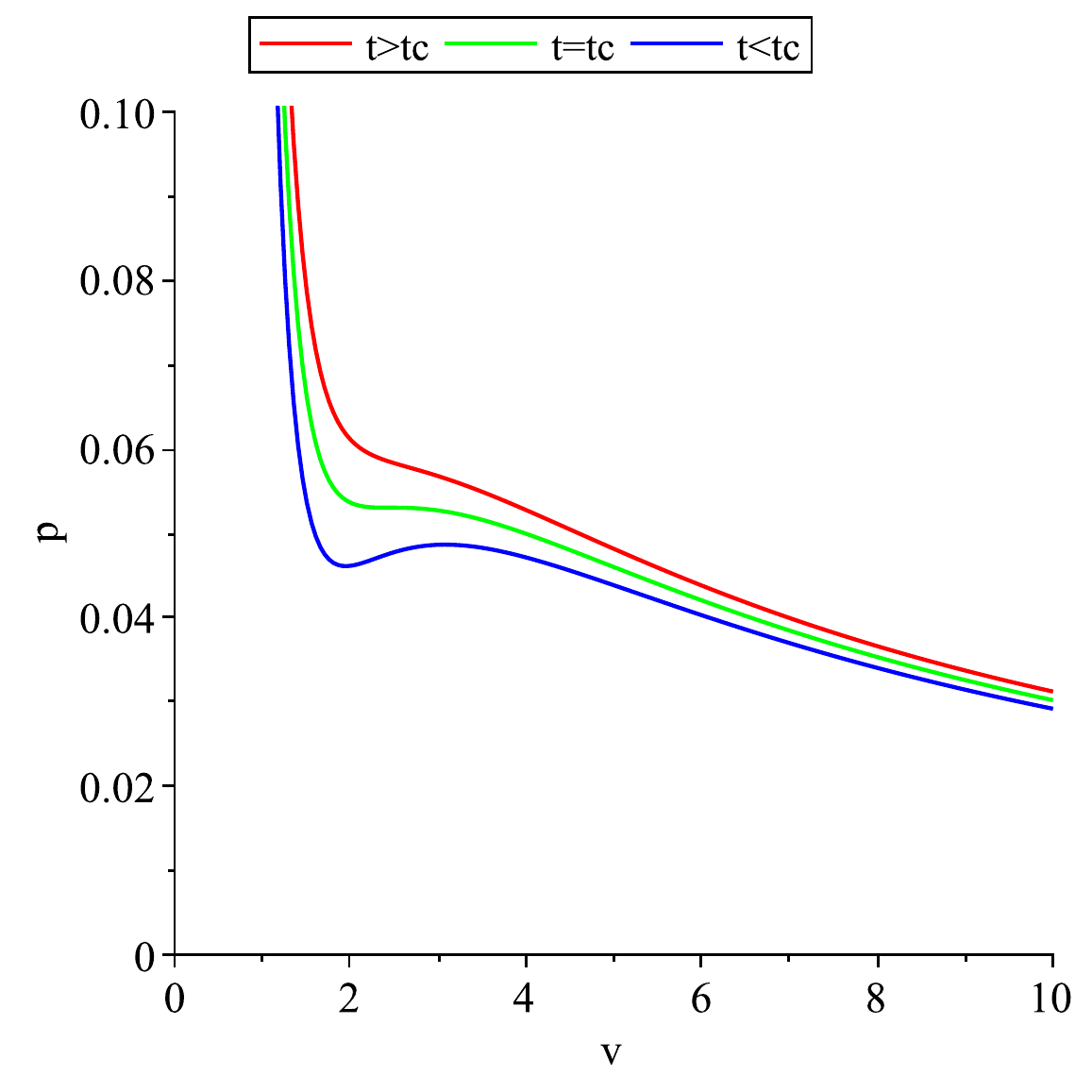}
    \end{subfigure}
    \begin{subfigure}{0.29\textwidth}
    \centering
    \includegraphics[width=\textwidth]{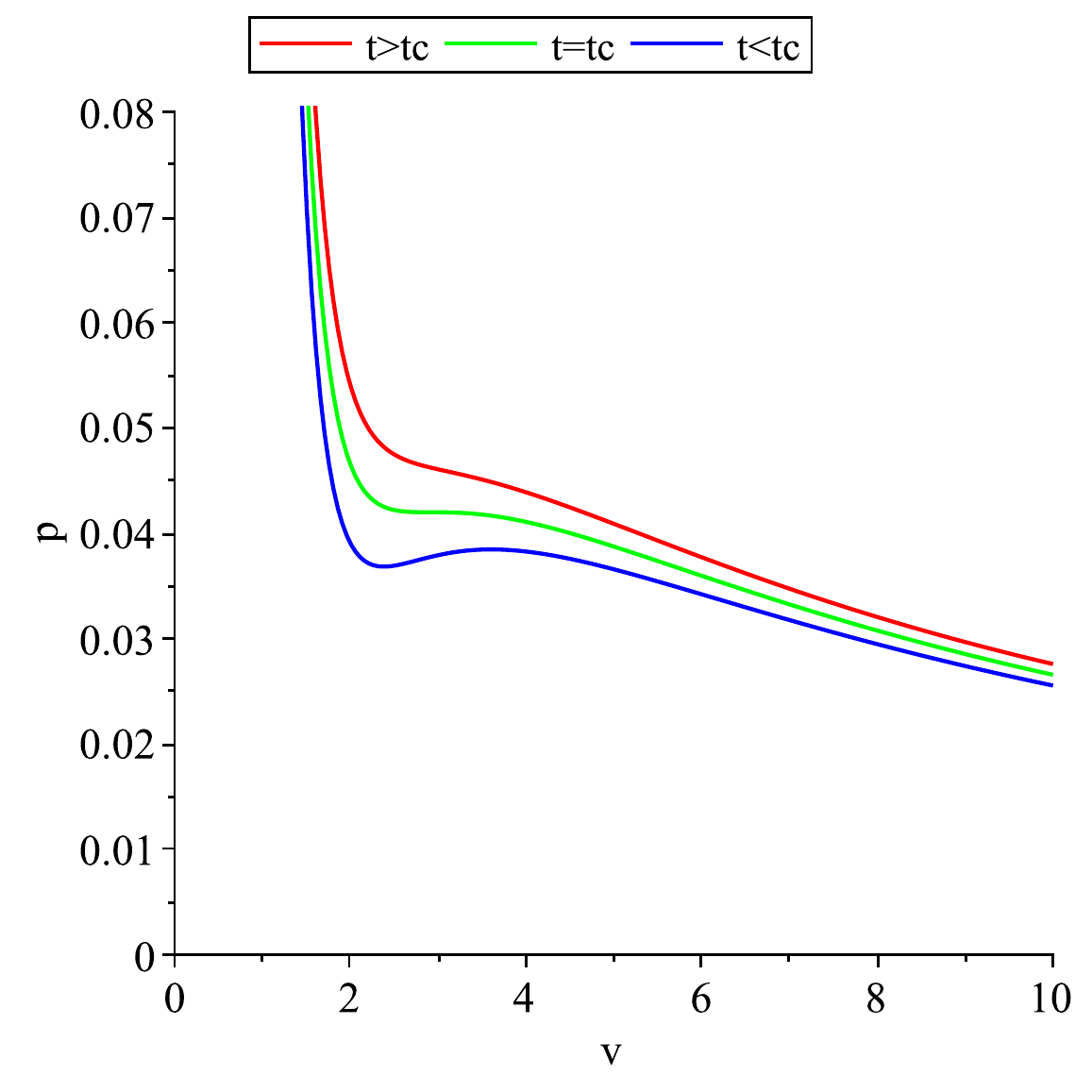}
    \end{subfigure}
    
    \begin{subfigure}{0.29\textwidth}
    \centering
    \includegraphics[width=\textwidth]{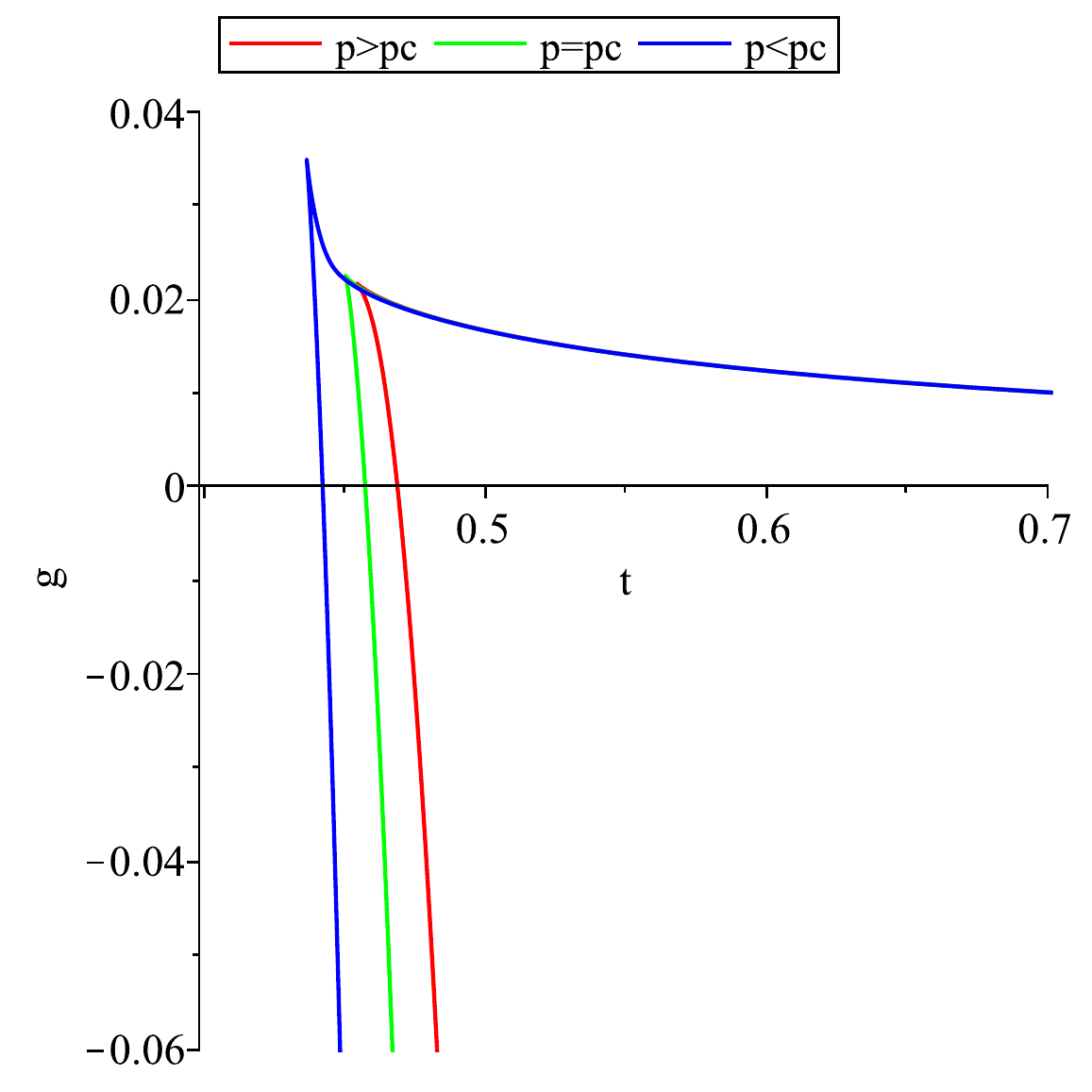}
    \caption{$b_{2}=1$}
    \end{subfigure}
    \begin{subfigure}{0.29\textwidth}
    \centering
    \includegraphics[width=\textwidth]{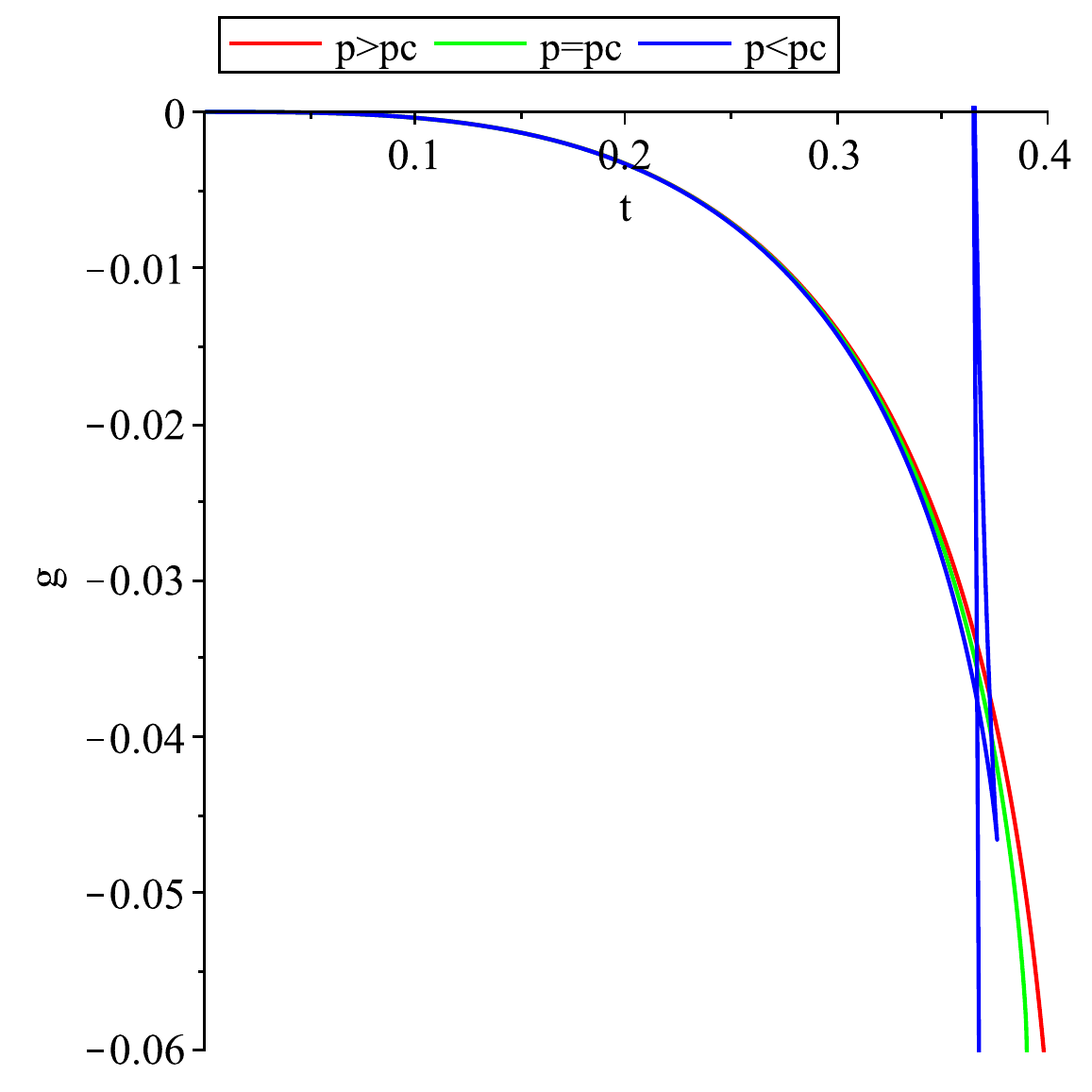}
    \caption{$b_{2}=0$}
    \end{subfigure}
    \begin{subfigure}{0.29\textwidth}
    \centering
    \includegraphics[width=\textwidth]{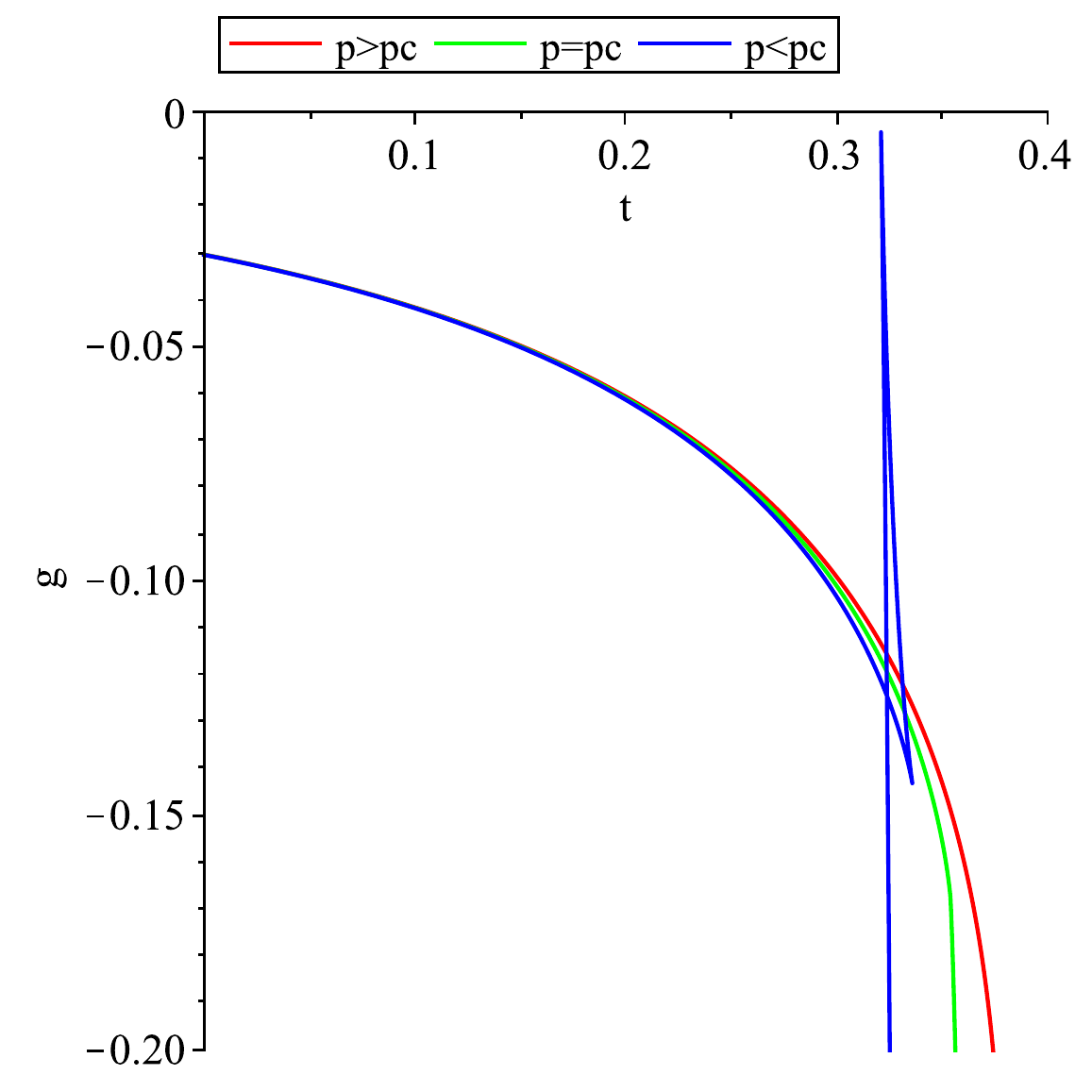}
    \caption{$b_{2}=-1$}
    \end{subfigure}
    \caption{\textbf{ Phase Behaviour for $d=6$, $q=0$, $b_{1}=1$ black holes.} \textit{Top}: Three $p-v$ diagrams for varying values of $b_{2}$ at constant temperature slices. \textit{Bottom}: Corresponding $g-t$ plots of constant pressure with $b_{2}$ values displayed below.The images on the left display critical temperature/pressure pressures for a non-stable black hole.}
    \label{6dpvgt0q}
\end{figure}

To examine the phase behaviour of the uncharged solutions, we first  set $b_{1}=1$, with the results displayed in Figure~\ref{6dpvgt0q}. For $b_{2}=1$, corresponding to the standard horizon geometries \cite{Frassino_2014},  
we get a cusp structure for the  $g-t$ diagram, and a maximal pressure in the $p-v$ diagram. There is also a minimum, temperature-dependent volume for which $p=0$.
As in  the 5 dimensional case, the black hole will undergo a Hawking/Page transition from a large black hole into thermal AdS. 

Things become more interesting as the value of $b_2$ changes.  As $b_2$ decreases, we 
 recover standard Van der Waals behaviour with a single oscillation in the $p-v$ diagram and the familiar swallowtail structure in the Gibbs free energy diagram.

Unlike the 5 dimensional case, this intersection of the swallowtail 
occurs below the $g=0$ axis and therefore is a genuine  first order large/small first order phase transition between  stable black holes -- this transition is generally not observed for uncharged black holes. 
As $b_2$ becomes negative, no further qualitative changes in phase behaviour are seen.  

\begin{figure}[H]
    \centering
    \begin{subfigure}{0.35\textwidth}
    \centering
    \includegraphics[width=\textwidth]{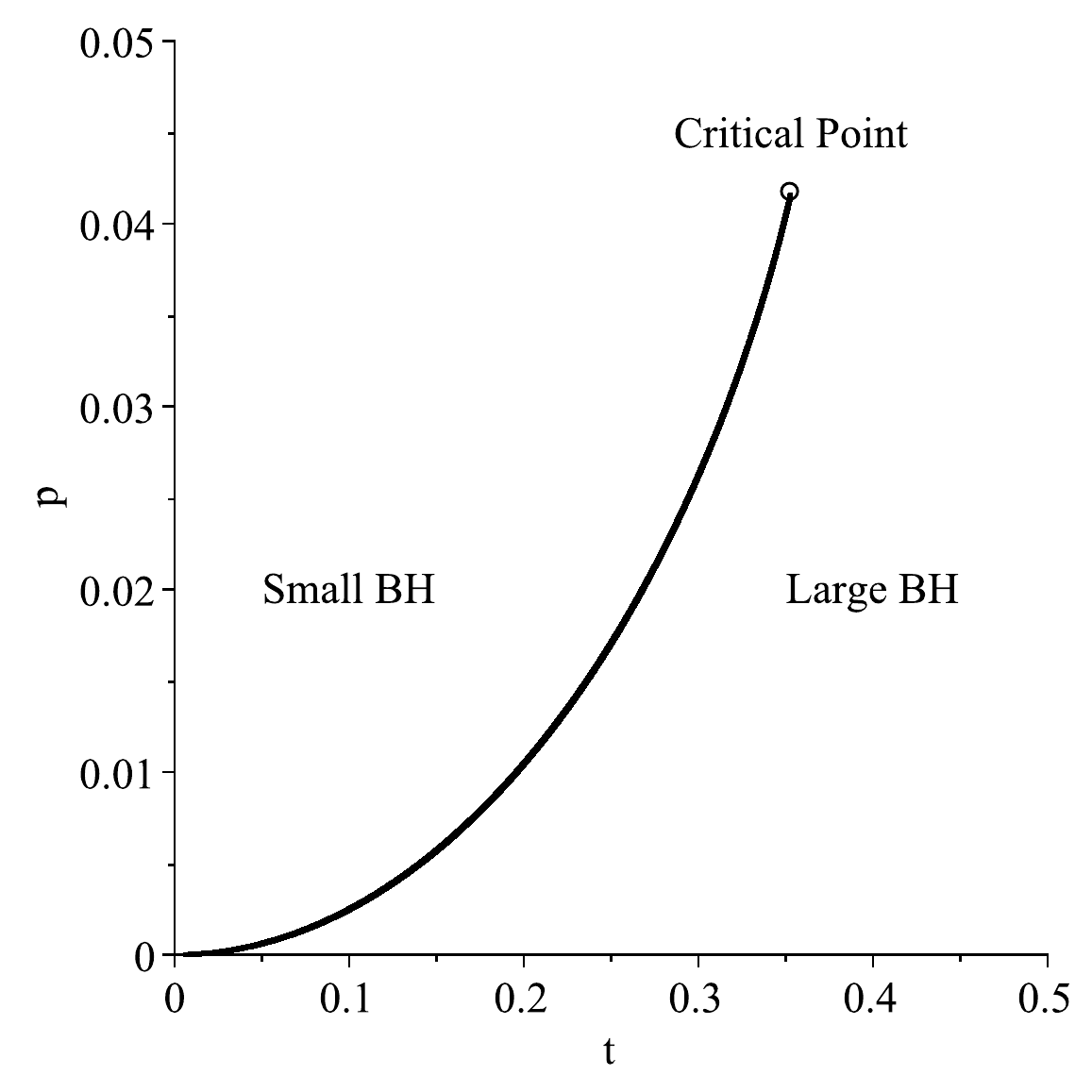}
    \end{subfigure}
    \begin{subfigure}{0.35\textwidth}
    \centering
    \includegraphics[width=\textwidth]{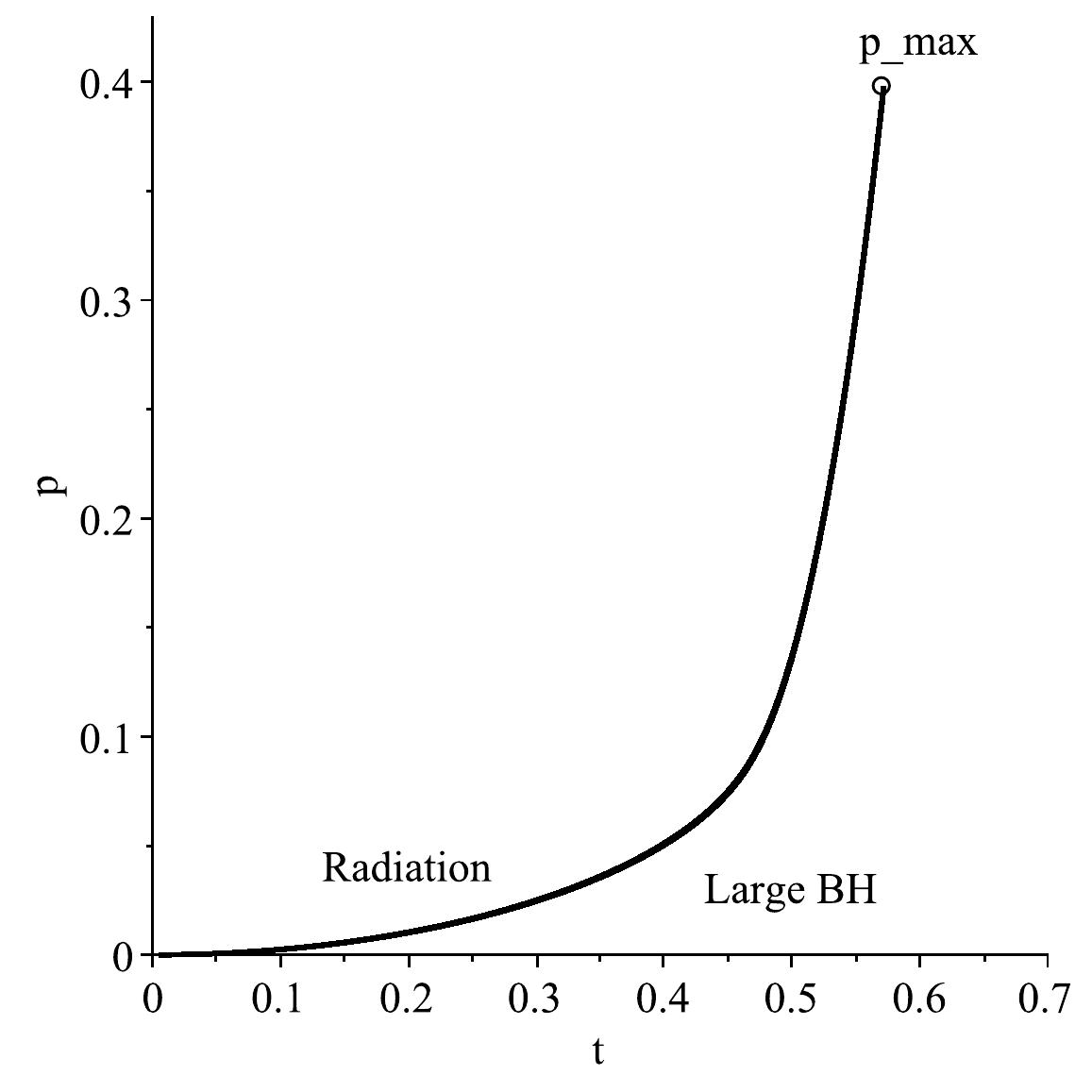}
    \end{subfigure}
    \caption{\textbf{Coexistence curves for $d=6$,   $q=0$,\, $b_{1}=1$ black holes.} \textit{Left:} $p-t$ phase diagram for $b_{2}=-1$ displaying first order transition between Small/Large BH, which terminates at the critical point. \textit{Right:} $p-t$ phase diagram for $b_{2}=1$ here we have a transition between Large BH and AdS radiation which terminates at the maximum pressure.
    }
    \label{pt6duncharged}
\end{figure}

The phase diagrams are depicted in Figure~\ref{pt6duncharged}. Here we can see that when we have $b_{2}=1$ we observe a Hawking/Page transition between a large black hole and thermal AdS, whereas for $b_{2}=-1$ we have the standard Van der Waals transition from a large black hole  to a small one.

We can notice something between the phase diagram displayed on the right of Figure~\ref{pt6duncharged} and that of the right of Figure~\ref{5DUncharged1}. In the 5 dimensional case there is a sharp but smooth bend in the diagram which corresponds to the critical point of the unstable black hole;  however in the 6 dimensional case the change after this point is not as dramatic.  This  can be easily explained with reference to the $g-t$ diagram. In the 6 dimensional case there is  no swallowtail behaviour for $b_{1}=b_{2}=1$; instead we only observe a cusp. The coexistence curve is correspondingly less sharp, although we still observe a steep slope as $p_{max}$ is approached. In the $g-t$ diagram, the curves for $p_{c}$ (the only  value that is a solution to \eqref{criticalpointsequation}) and $p_{max}$ are close together, with  only a small difference in temperature between them.

\begin{figure}[H]
    \centering
    \begin{subfigure}{0.4\textwidth}
    \centering
    \includegraphics[width=\textwidth]{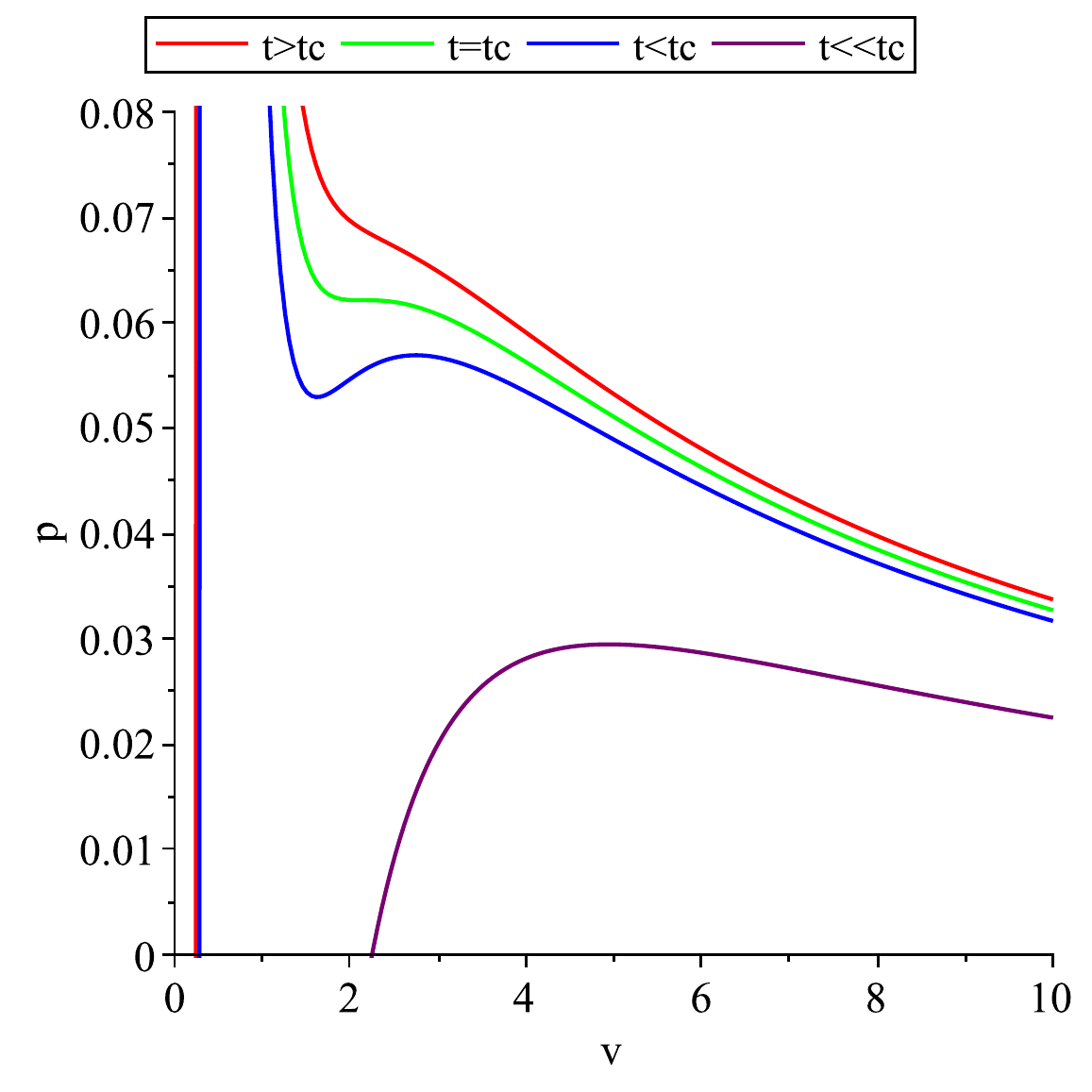}
    \end{subfigure}
    \begin{subfigure}{0.4\textwidth}
    \centering
    \includegraphics[width=\textwidth]{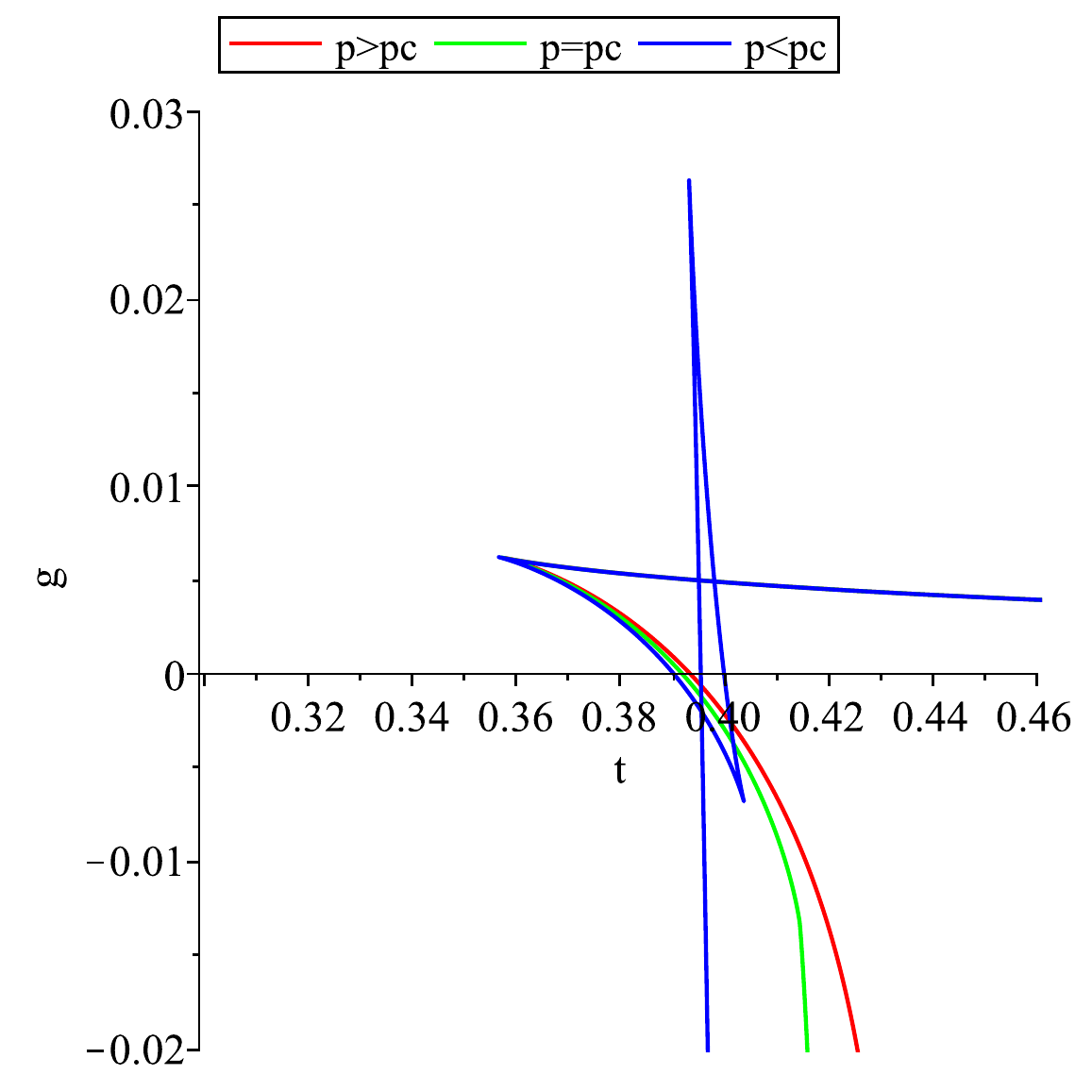}
    \end{subfigure}
    \caption{\textbf{Phase Behaviour for $d=6$, $q=0$, $b_{1}=1$, $b_{2}=0.5$ black holes.} \textit{Left:} $p-v$ diagram for constant temperature slices dislaying Van der Waals oscillations. \textit{Right:} $g-t$ diagram for constant pressure slices.}
    \label{6Dq0pvgtb205}
\end{figure}

\begin{figure}[H]
    \centering
    \includegraphics[width=0.4\textwidth]{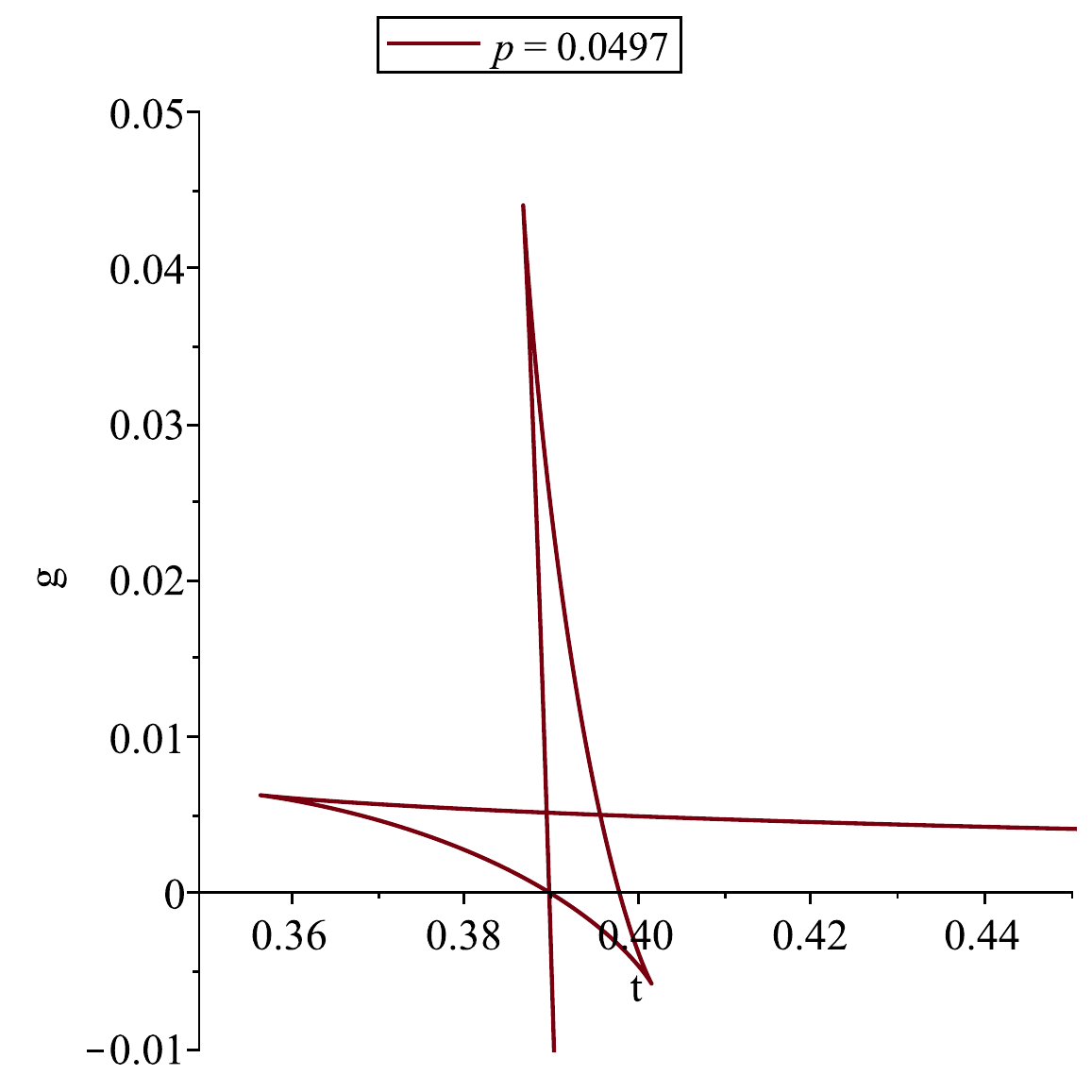}
    \caption{\textbf{Novel Triple Point for $d=6$, $b_{1}=1$,$b_{2}=0.5$}. Gibbs temperature diagram showing the two-branch intersection occuring along the $g=0$ axis.}
    \label{6DNewTrip}
\end{figure}

\begin{figure}[H]
    \centering
    \includegraphics[width=0.5\textwidth]{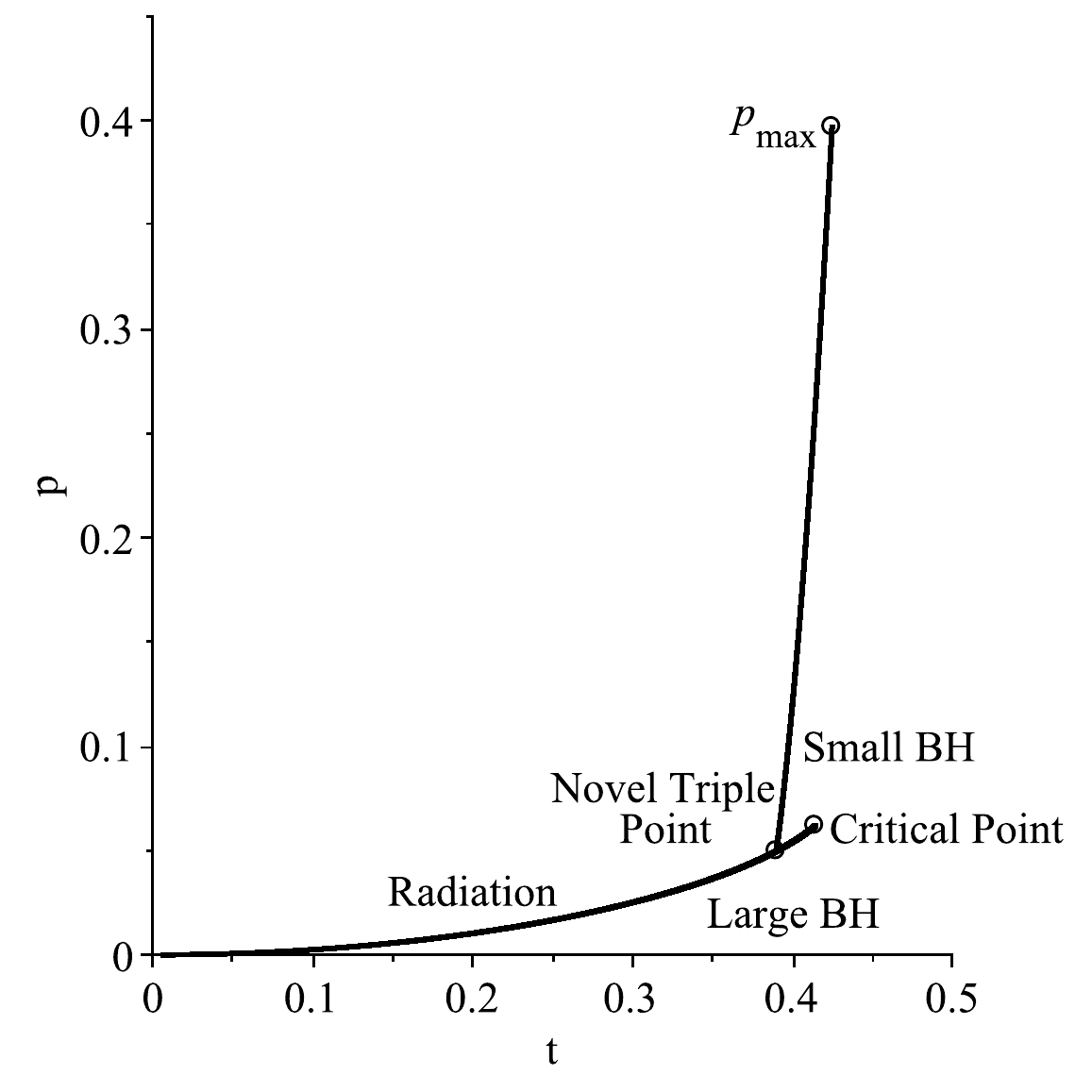}
    \caption{\textbf{ Coexistence Curves for $d=6$, $b_{1}=1$ , $b_{2}=0.5$, $q=0$ black holes.}  The $p-t$ phase transition diagram displays the novel triple point. The large/small BH transition terminates at the critical point while the small/radiation coexistence line extends up to the maximum pressure.
    }
    \label{ptnoveltrip}
\end{figure}

For $1 >b_2 > 0$ swallowtail behaviour occurs and the possibility of new phenomenon emerges -- that of a new kind of black hole triple point.   In  Figure~\ref{6Dq0pvgtb205} we see
(for $b_{1}=1$ and $b_{2}=0.5$) that for $p<p_c$ we can have 
 a large black hole undergoing a first order phase transition into a small black hole that in turn undergoes a transition to thermal AdS as the temperature is lowered further.   If we decrease the pressure even further, we  arrive at Figure~\ref{6DNewTrip}, in which we have the large/small transition occurring on the $g=0$ axis. This implies a novel triple point where we have the coexistence  large and small black holes with thermal AdS. 
 The phase diagram displaying this novel triple point can be found in Figure~\ref{ptnoveltrip}. For sufficiently low pressures, there are only 2 phases, thermal AdS and the large black hole.  As the pressure increases, the triple point emerges where the small black hole phase coexists with the other two.  At  pressures above the novel triple point pressure we observe the three distinct phases as the temperature varies. There is a further critical pressure at which small and large black holes are no longer distinct phases; above this pressure we again have just a single Hawking/Page transtion between thermal AdS and a black hole.

Although within the range of $0 < b_{2} <1$   swallowtail behaviour is observed, not all values of $b_2$ in that range yield a novel-triple point. For $b_{1}=1$ we find that  only the range $0<b_{2}<\frac{2}{3}$ for which the novel triple point occurs. This can be obtained by solving for the critical volume and temperature, in turn yielding the constraint \begin{equation}
    \frac{\left(-3 b_{2}+12\right) \sqrt{b_{2}^{2}-10 b_{2}+9}-3 b_{2}^{2}+31 b_{2}-36}{\pi\left(\sqrt{b_{2}^{2}-10 b_{2}+9}+b_{2}-3\right)^{2}\left(b_{2}-9+\sqrt{b_{2}^{2}-10 b_{2}+9}\right)} < 0
\end{equation}
that must be satisfied in order to obtain the critical pressure; its solutions are 
$0<b_{2}<\frac{2}{3}$. 
If $b_{2}=2/3$, we only have one degenerate critical pressure, at which the novel triple point   terminates at the critical point of the black hole, pushing the small branch out and only leaving a large/radiation transition.

\begin{figure}[H]
    \centering
    \begin{subfigure}{0.4\textwidth}
    \centering
    \includegraphics[width=\textwidth]{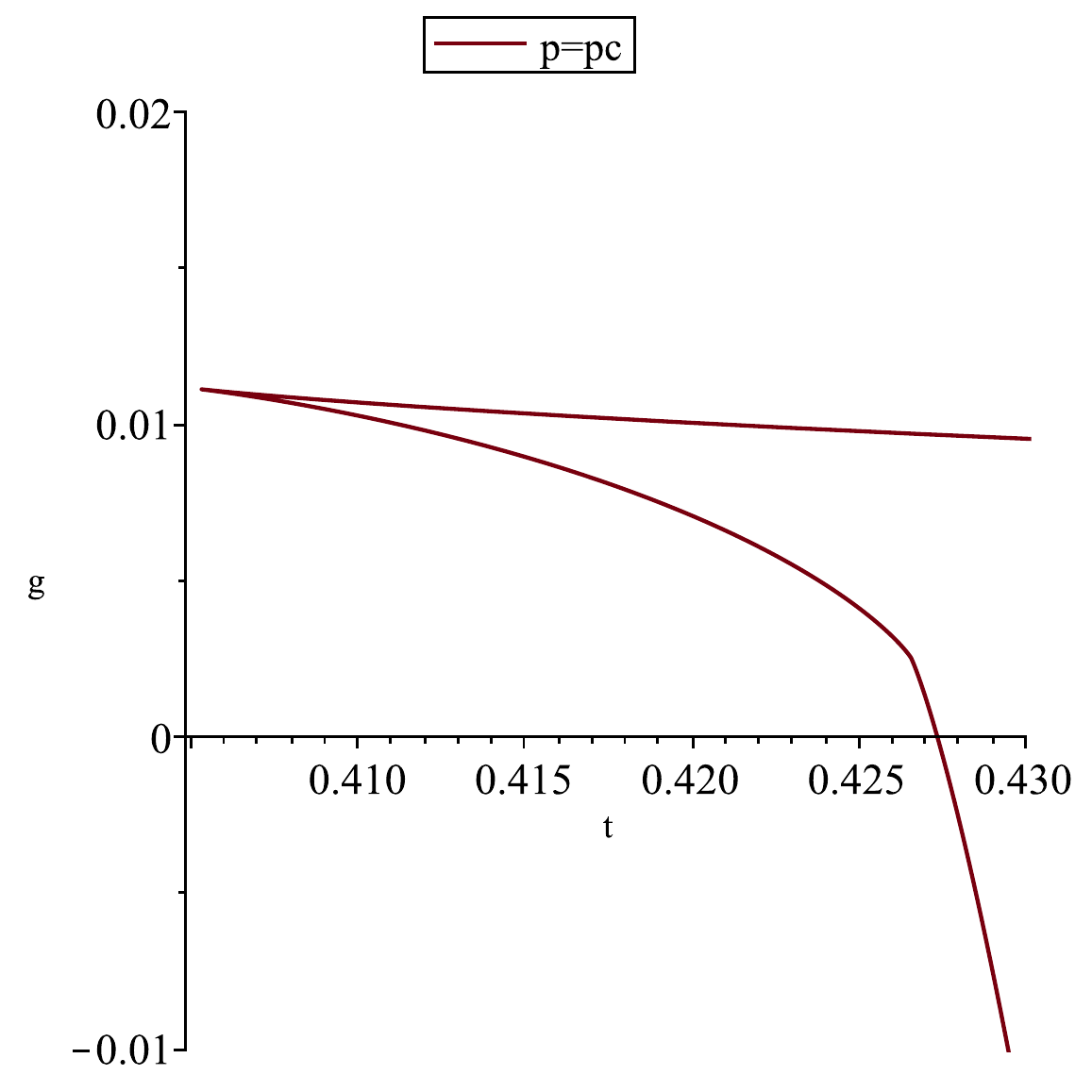}
    \caption{$b_{2}=0.7$}
    \end{subfigure}
    \begin{subfigure}{0.4\textwidth}
    \centering
    \includegraphics[width=\textwidth]{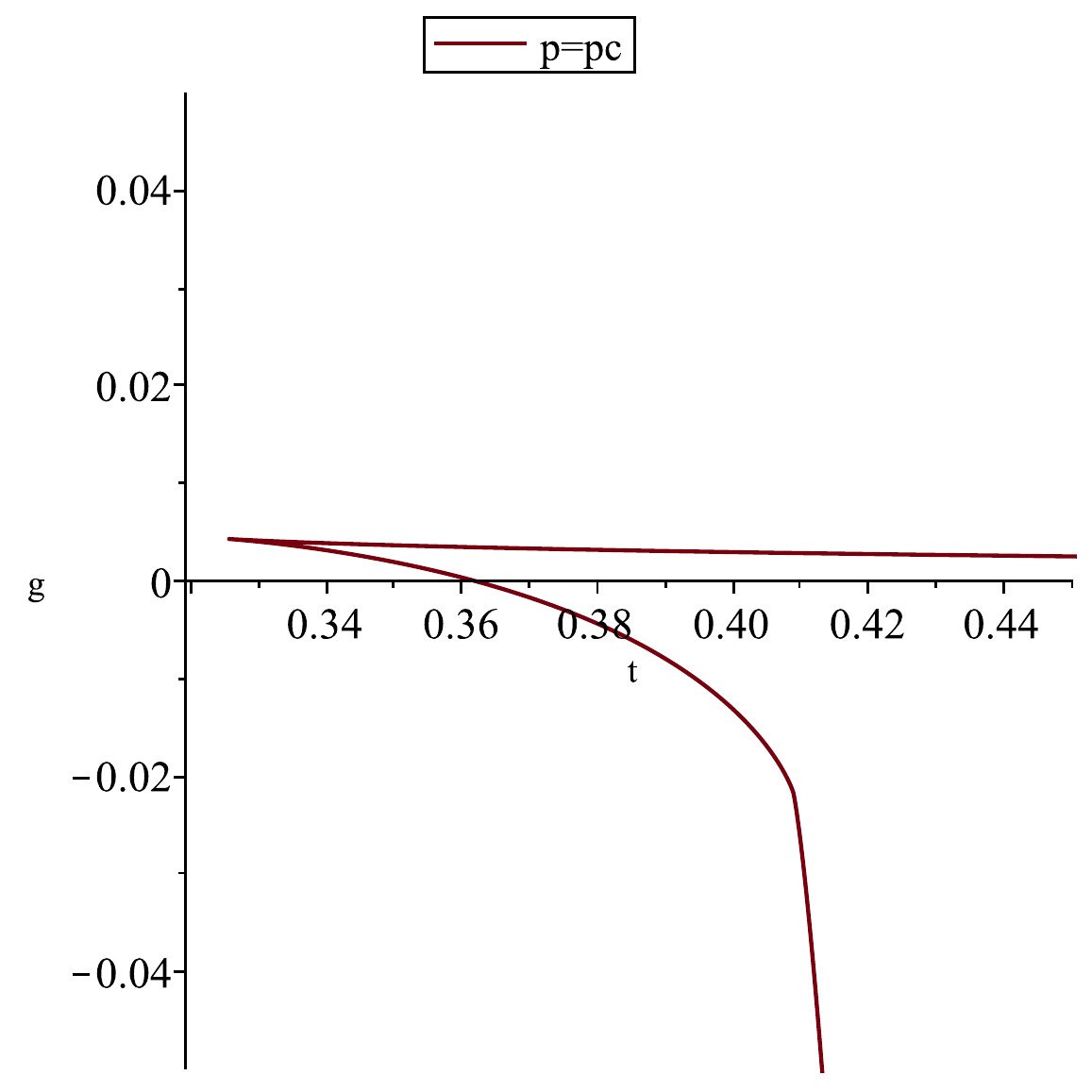}
    \caption{$b_{2}=0.4$}
    \end{subfigure}
    \caption{Comparing the two $g-t$ diagrams with two different values of $b_{2}$ and showing the position of the sharp "Corner" being above and below the $g=0$ axis.}
    \label{gtcompare}
\end{figure}

While the above conditions provide the information needed to know if a triple point is possible, we can also gain information on this from  the $g-t$ diagram. In Figure~\ref{gtcompare} we see that moving rightward from the cusp on the lower branch, there is a discontinuity in the first derivative of $g$, indicative of a small/large critical point.   If this point  is above the $g=0$ axis no triple point will occur, whereas if it is below then we find the novel triple point. If this point intersects the $g=0$ axis then the novel triple point merges with the small/large critical point.

Turning to $b_1 < 0$, we must have either $b_2<0$ or $b_2 \geq 9b_1^2$ 
in order to have a real and positive critical volume. 
However the latter case yields a negative critical temperature, and so critical behaviour can take place only for $b_2<0$.  The only phase behaviour we observe in the range $9b_1^2> b_2 > 0$ is that of a Hawking-Page transition.

We close by noting that  throughout this subsection we have chosen  
$b_{2}-b_{1} \leq 0$. Hence none of the black holes we consider will   possess a vacuum singularity outside  the origin.  

%While it is possible to observe all of our phenomena displayed above for differing values of $b_{1}$ and $b_{2}$ it is unnecessary to display any further. \tcr{Maybe a future paper we can examine solutions who will posses $b_{2}-b_{1}^2 >0$ and will give us vacuum singularities.} 

\subsubsection{Charged ELBHs}

Including charge, the equation of state is now 
\begin{equation}
p=\frac{t}{v}-\frac{3 b_{1}}{\pi v^{2}}+\frac{2 t b_{1}}{v^{3}}+\frac{q^{2}}{v^{8}}-\frac{b_{2}}{\pi v^{4}}
\end{equation}
with Gibbs free energy
\begin{equation}
    g=-\frac{\left(\frac{1}{4} v^{4}+b_{1} v^{2}\right)\left(\pi p v^{2}+3 b_{1}+\frac{b_{2}}{v^{2}}\right)}{4 \pi v\left(1+\frac{2 b_{1}}{v^{2}}\right)}+\frac{\frac{1}{5} \pi p v^{5}+b_{1} v^{3}+b_{2} v}{4 \pi}+\frac{q^{2}\left(14 v^{2}+40 b_{1}\right)}{96\left(v^{2}+2 b_{1}\right) v^{3}}\; .
\end{equation}
The critical temperature and critical volume relations \eqref{critt} and \eqref{critv} are now
\begin{equation}\label{ctemp6}
    t_{c}=\frac{6 b_{1} v_{c}^{6}+4 b_{2} v_{c}^{4}-8 \pi q^{2}}{\pi v_{c}^{5}\left(v_{c}^{2}+6 b_{1}\right)}
\end{equation}
\begin{equation}\label{cvol6}
    6 v_{c}^{8} b_{1}-\left(36 b_{1}^{2}-12 b_{2}\right) v_{c}^{6}+24 v_{c}^{4} b_{1} b_{2}+\left(-56 \pi v_{c}^{2}-240 \pi b_{1}\right) q^{2}=0.
\end{equation}
This latter equation is a quartic polynomial in $v^2_c$; analytic solutions can be obtained, but they are cumbersome and so we will not display them.

If $b_{1}=b_{2}=1$, which corresponds the constant curvature case \cite{Frassino_2014}, standard Van der Waals behaviour is observed, with a large/small first order transition occurring for $q>0.1$. This  was seen in the 5 dimensional case as well, so we shall not   display any phase diagrams for this case. Only one critical point is present for $q \geq 0.1$.
  More interesting behaviour occurs for values of $q < 0.1$. We find that more then one critical volume/temperature/pressure is possible, leading to the existence of triple points, previously observed for charged black holes in $d=6$ Lovelock gravity \cite{Frassino_2014}.

\begin{figure}[H]
    \centering
    \begin{subfigure}{0.4\textwidth}
    \centering
    \includegraphics[width=\textwidth]{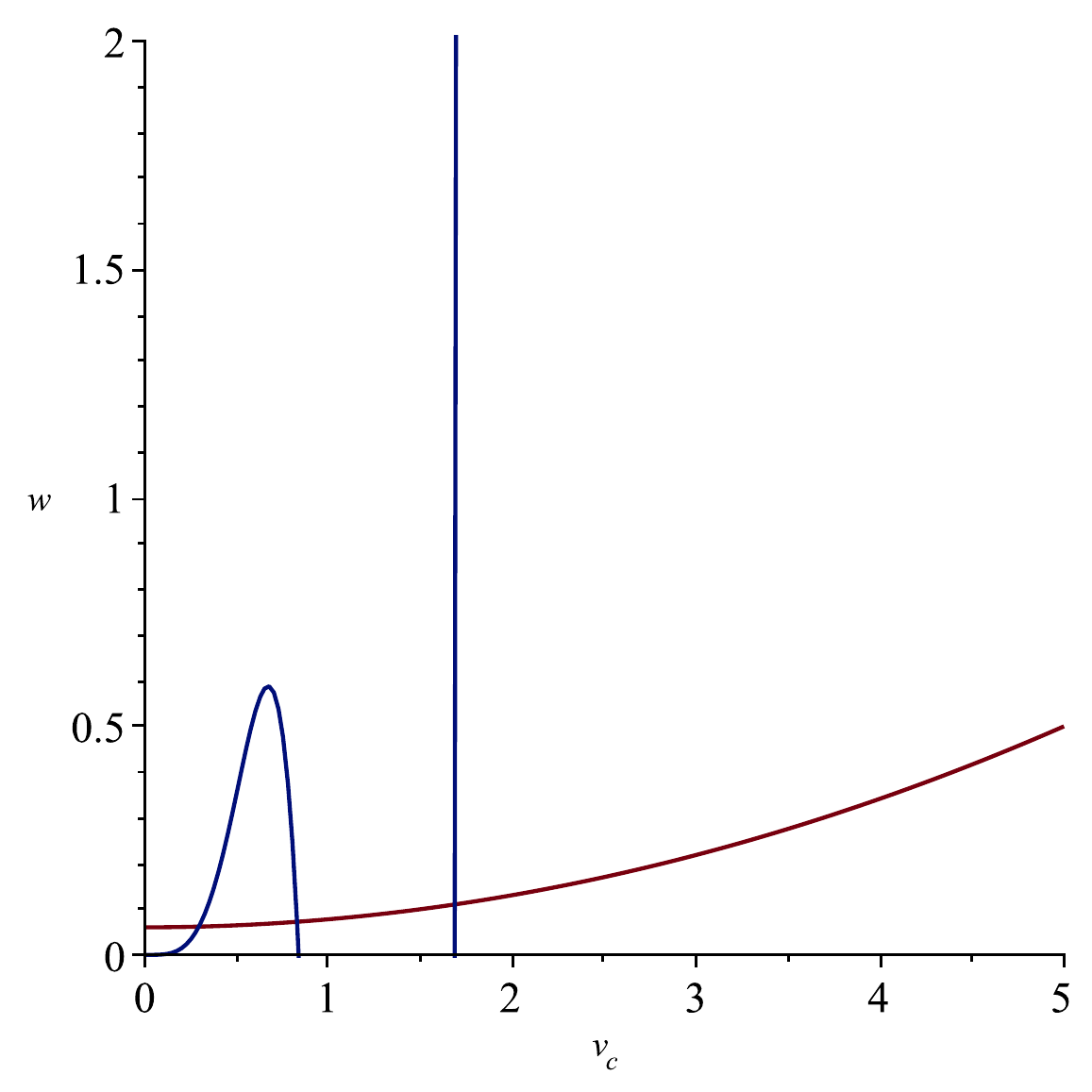}
    \caption{q=0.01}
    \end{subfigure}
    \begin{subfigure}{0.4\textwidth}
    \centering
    \includegraphics[width=\textwidth]{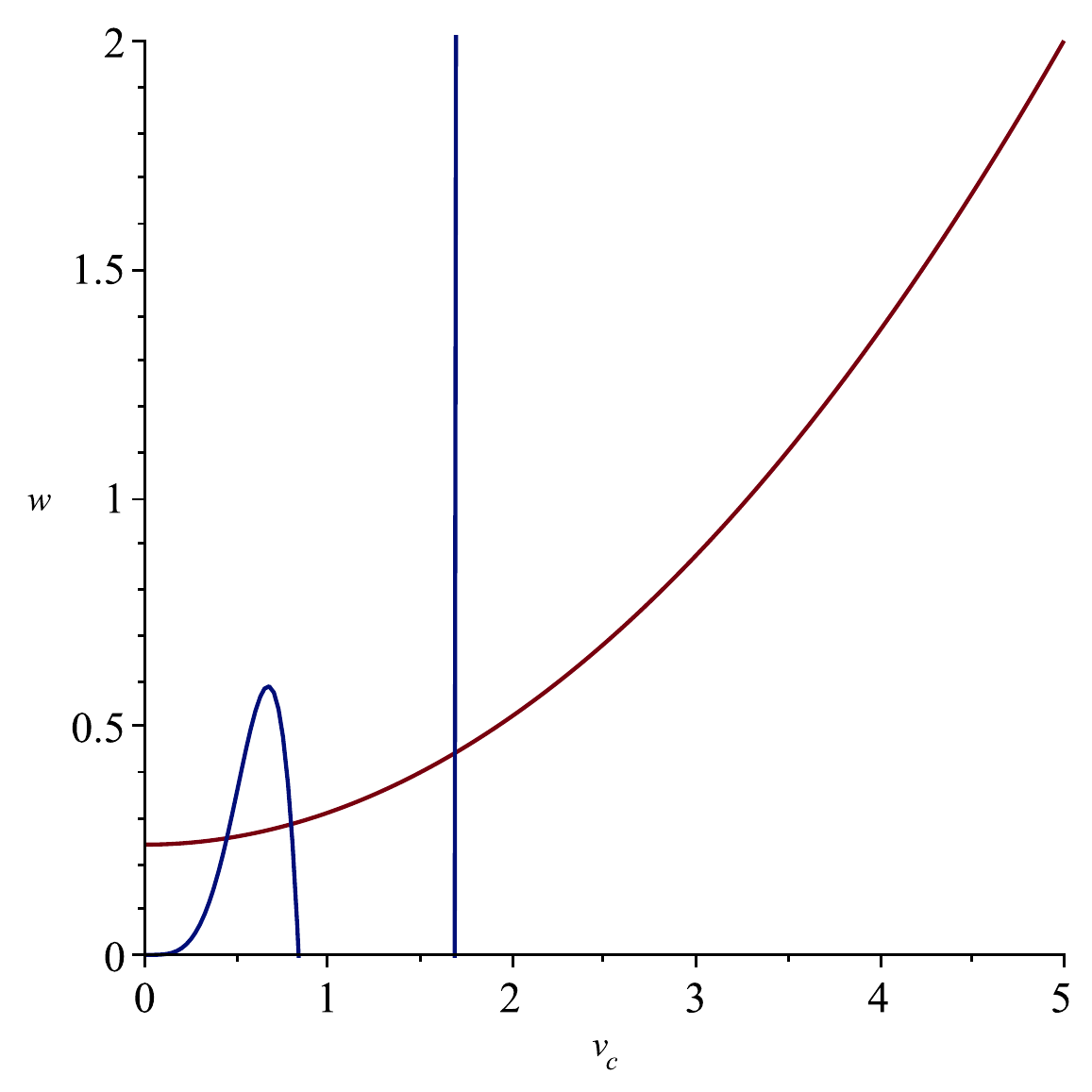}
    \caption{q=0.02}
    \end{subfigure}
    \caption{Plots of $w_{1}$(red) and $w_{2}$ (blue) in 6 dimensions for constant charge, showing three intersection between the two functions, giving the possibility of a triple point. Here $b_{1}=0.8$ and $b_{2}=0.5$.
    }
    \label{wplotconstq}
\end{figure}

\begin{figure}[H]
    \centering
    \begin{subfigure}{0.45\textwidth}
    \centering
    \includegraphics[width=\textwidth]{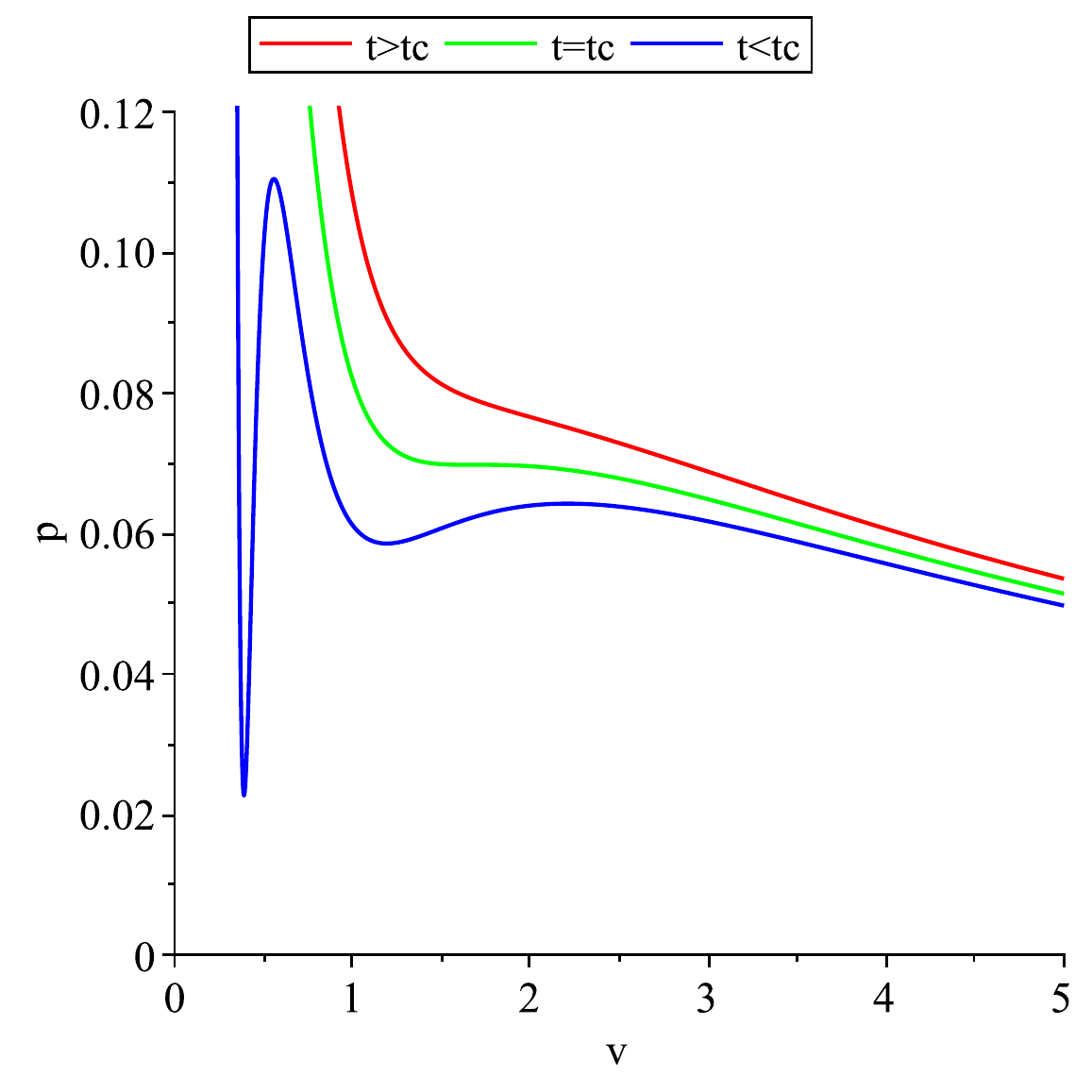}
    \end{subfigure}
    
    \begin{subfigure}{0.32\textwidth}
    \centering
    \includegraphics[width=\textwidth]{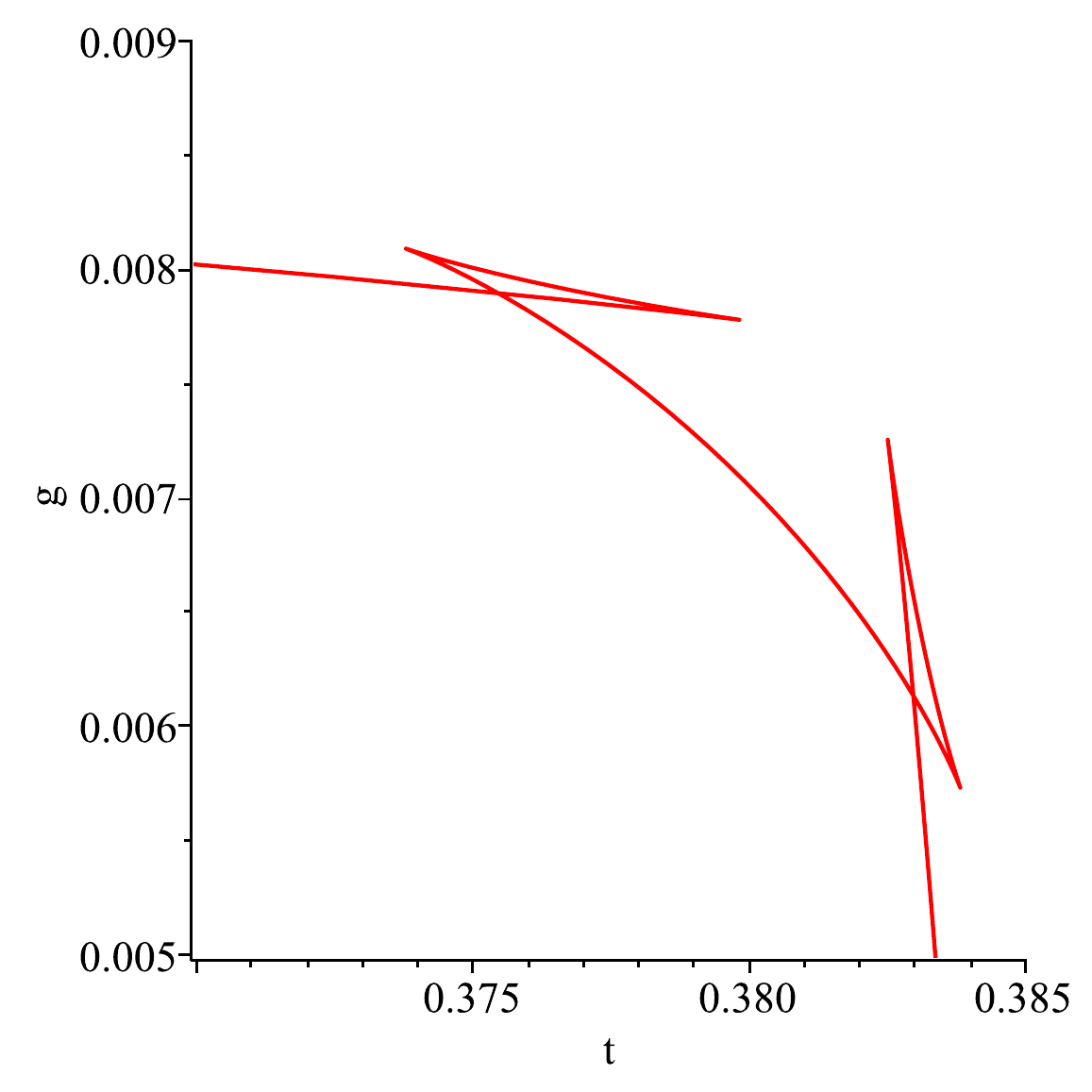}
    \caption{p=0.06687}
    \end{subfigure}
    \begin{subfigure}{0.32\textwidth}
    \centering
    \includegraphics[width=\textwidth]{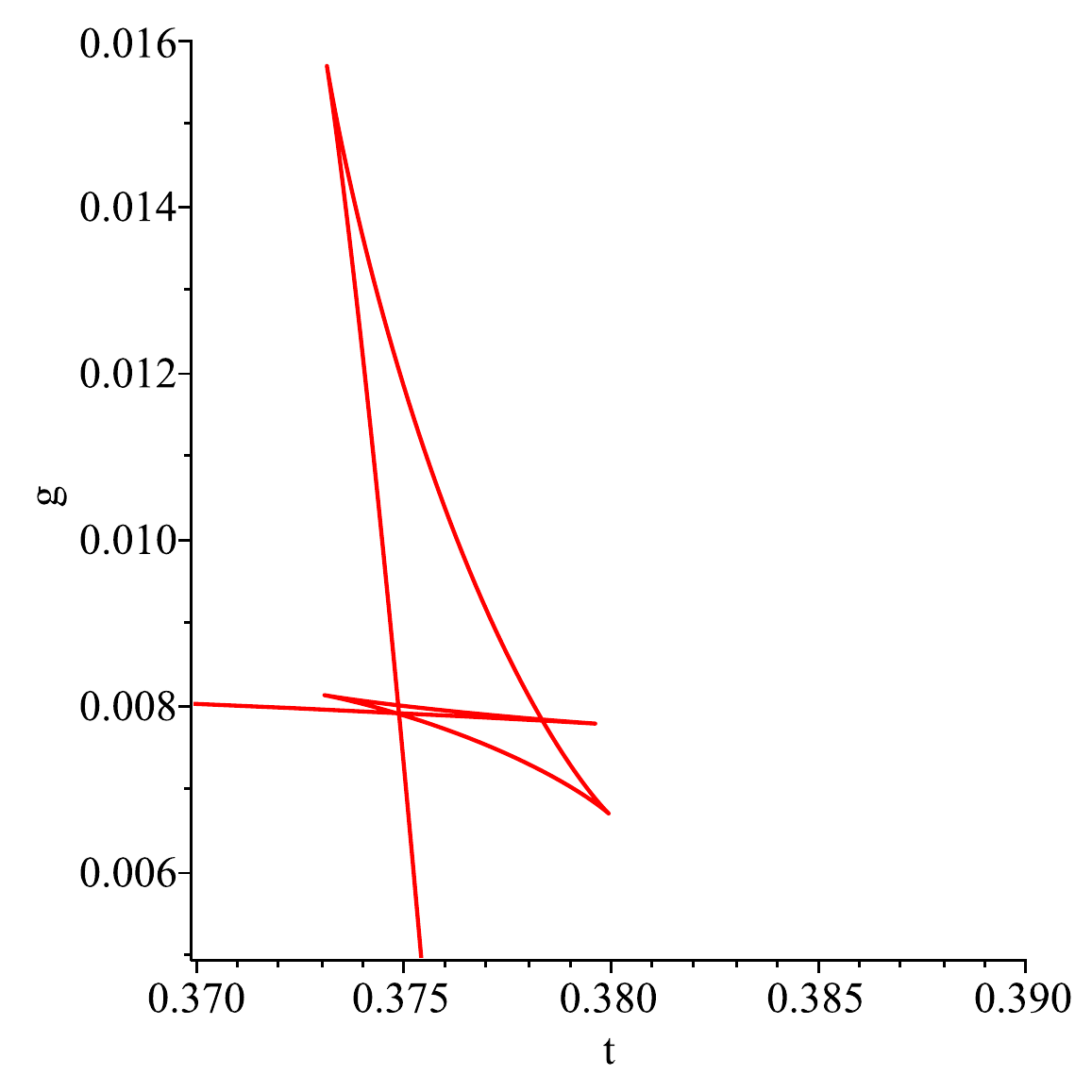}
    \caption{p=0.06127}
    \end{subfigure}
    \begin{subfigure}{0.32\textwidth}
    \centering
    \includegraphics[width=\textwidth]{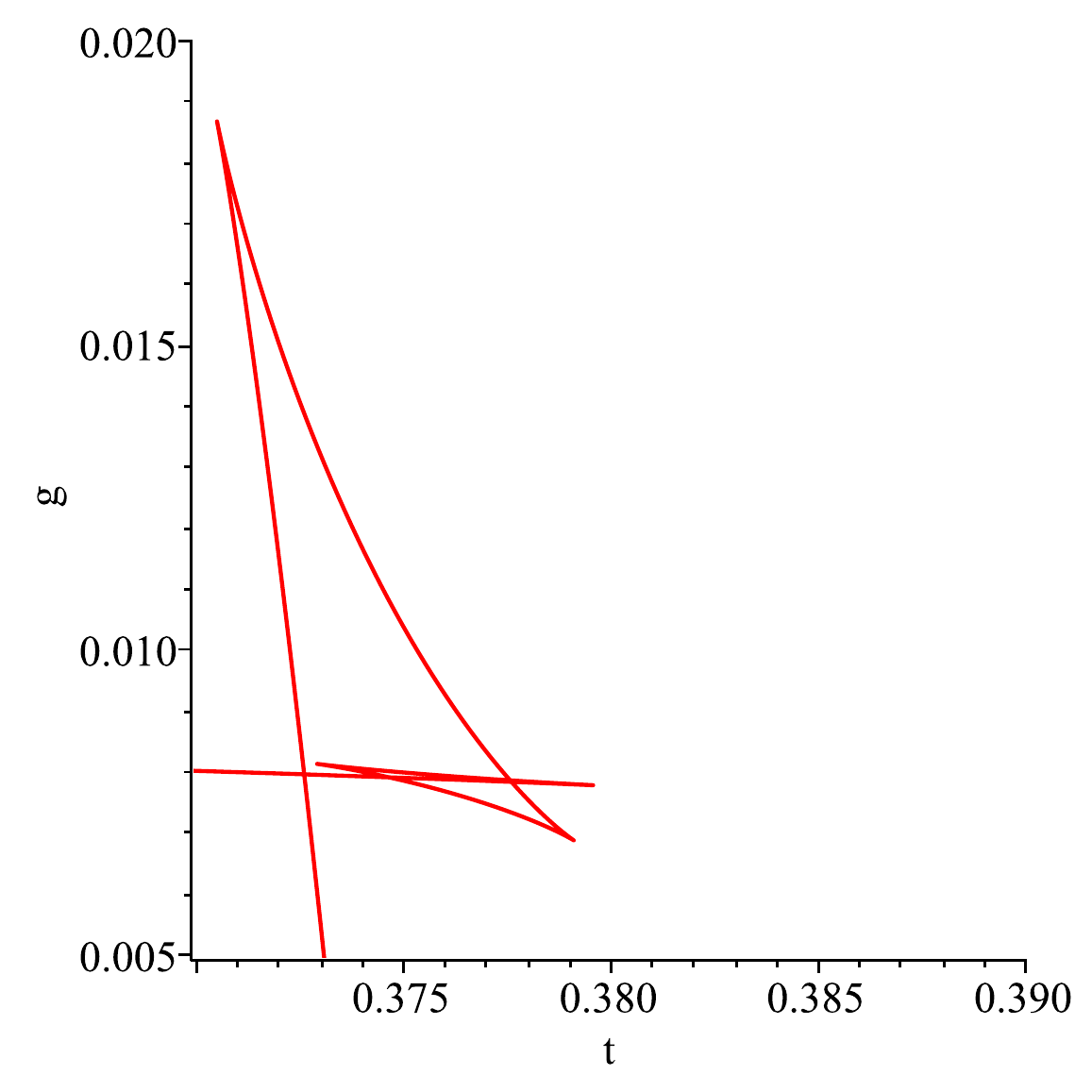}
    \caption{p=0.05987}
    \end{subfigure}
    \caption{{\bf Phase Behaviour for $d=6$, $q=0.02$,$b_{1}=0.8$,$b_{2}=0.5$ black holes.}   {Top:} The $p-v$ diagram for three constant temperature slices located around the tri-critical temperature. We can see in the blue line 2 oscillations, similar to the constant curvature case.  {Bottom:} Three $g-t$ plots for constant pressure slices are shown. As we decrease the pressure from left to right we see the presence of two swallow tails, which eventually intersect, then separate again.}
    \label{mytriplepoint}
\end{figure}

\begin{figure}[H]
    \centering
    \includegraphics[width=0.5\textwidth]{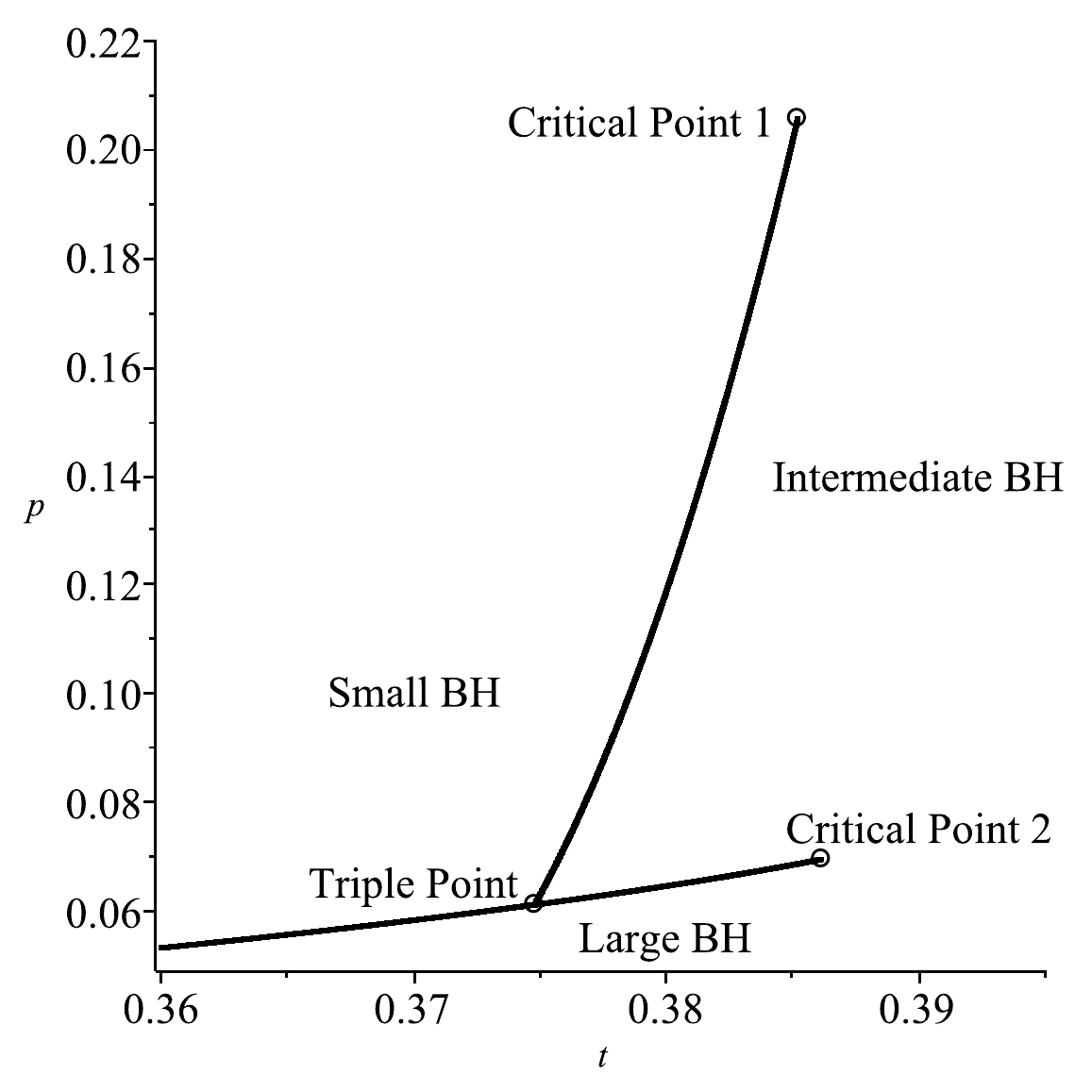}
    \caption{{\bf Coexistence Curves for $d=6$,  $b_{1}=0.8$, $b_{2}=0.5$,$q=0.02$ black holes.} We observe the triple point at the intersection of the curves. }
    \label{ptmytrip}
\end{figure}

 To see what happens to the triple point if the horizon curvature is not constant, it is useful to rewrite the critical volume equation \eqref{cvol6} as
\begin{equation}
   w_1 \equiv \left(56 \pi v_{c}^{2}+240 \pi b_{1}\right) q^{2}=6 v_{c}^{8} b_{1}-\left(36 b_{1}^{2}-12 b_{2}\right) v_{c}^{6}+24 v_{c}^{4} b_{1} b_{2} \equiv w_2
\end{equation}
and search for the intersection points.  A necessary condition for a triple point to occur is that
there are three intersection points for $v_c >0$. This will indeed occur as long as the signs of the coefficients in $w_2$ alternate -- the rule of signs then indicates there are three positive real roots for $v_c^2$ (and hence for $v_c$), in turn implying two distinct regions where $w_2>0$ for $v_c>0$, one of which has a maximum.  Since $w_1$ is a quadratic in $v_c$ with coefficient $q^2$, there will be three intersection points for sufficiently small $q>0$ and $b_1>0$. An example is given in
Figure~\ref{wplotconstq} for
 $b_{1}=0.8$ and $b_{2}=0.5$. The emergence of the triple point for increasing pressure is shown in Figure~\ref{mytriplepoint}, with the phase diagram given in Figure~\ref{ptmytrip}.

We find that for a given value of $b_1$, the triple point in Figure~\ref{ptmytrip} moves to the right as $b_2$ increases. For sufficiently large $b_2$, the triple point merges with the large/intermediate critical point. Conversely, fixing $b_2$ and increasing $b_1$ moves the triple point to the left.  This illustrates how changing the horizon geometry of exotic black holes modifies their phase behaviour.

\section{Summary \& Conclusion}

Our investigation of thermodynamic behaviour for exotic black holes has uncovered a number of interesting results. 

First, concerning the generality of our results, we note that the existence of a formal solution to the field equations  for arbitrary values of $b_1$ and $b_2$ does not ensure that a base manifold  satisfying
\eqref{Eul-Lov}  exists
for such values.  In general the existence of solutions to \eqref{Eul-Lov} for given $(b_1,b_2)$ is an open question. However it is possible to reverse the roles of our parameters and treat $b_{1}$ as a continuous parameter (easily incorporated by
adding in a global monopole) while keeping $b_{2}$ fixed, which is consistent with the approach considered in \cite{Dotti_2005}. In that case the Bohm $(p,q)_{2m}$ metric was found to be a base manifold
satisfying
\eqref{Eul-Lov}, and a set of allowed values of $b_2$ (denoted $\theta$ in \cite{Dotti_2005}) were obtained as a function of the integer $q$. The smallest value $q=2$ yields $b_2=12$. It is straightforward to show that values of  $4 < b_{1} < 4.24$ exhibit the  radiation/large black hole transition, and for  $b_{1}>4.24$  the novel triple point occurs. We expect that other base manifolds can likewise be found whose associated black holes display the features we have observed.

The most interesting is that of a novel triple point
between thermal AdS (radiation), and uncharged large  and small black holes in 6 dimensions. This phase behaviour was overlooked in previous studies \cite{Frassino_2014}, and arises as a consequence of the exotic geometry of the horizon.  We likewise observe a range of large/intermediate/small black hole
triple point behaviour in the charged case in $d=6$ as we adjust the parameters of the horizon geometry.

Another interesting result is the generalizations  \eqref{vacsol}    of massless topological black holes in Einstein gravity
\cite{Mann:1997jb,Aminneborg:1996iz,Smith:1997wx}. For these exotic Gauss-Bonnet black holes two horizons are possible, yielding a richer   set of possibilities warranting further study.  Negative mass solutions generalizing those in Einstein gravity \cite{Mann:1997jb,Smith:1997wx} are also possible. We leave a more detailed study of these object for future investigations.

A study of 3rd order Lovelock gravity, with the possibility of finding a quadruple point, would be interesting. There are two possibilities for a quadruple point. One is that of a novel uncharged quadruple point where we have two swallowtails intersecting each other on the $g=0$ axis, giving large/intermediate/small/radiation coexistence point. Another would be that in the charged case, where four black holes of distinct size merge at a single point in the phase diagram.   

More ambitious endeavours include promoting the topological parameter to a thermodynamic variable itself, generalizations to  to de Sitter spacetime, and obtaining rotating solutions. Work on these areas is in progress.
 
%Lastly, even though in early debate, the possibility of looking at these black holes in the novel 4D Gauss-Bonnet Lovelock gravity. 

\section*{Acknowledgements}
This work was supported in part by the Natural Sciences and Engineering Research Council of Canada.  We are grateful to Sourya Ray for a number of helpful discussions.
 
\section*{Note added}
As we were completing this paper we became aware of a similar study in 3rd order Lovelock gravity \cite{Farhangkhah:2021tzq}. This study considers only uncharged black holes.  We do not agree with some of their findings, particularly the failure to notice the presence of Hawking-Page transitions when relevant.

\bibliographystyle{unsrt}
\addcontentsline{toc}{section}{References}
\bibliography{References}

\end{document}